\documentclass[onecollarge,natbib]{svjour2}
\bibpunct{[}{]}{,}{n}{}{,} 
\smartqed  
\usepackage{graphicx}
\usepackage{subfig}

 \usepackage{mathptmx}   

\newcommand{\beq}{\begin{eqnarray}}
\newcommand{\eeq}{\end{eqnarray}}
\newcommand{\be}{\begin{eqnarray*}}
\newcommand{\ee}{\end{eqnarray*}}

\newcommand{\ie}{{\it i.e.}}

\newcommand{\cf}[1]{{Fig.~\ref{#1}}}

\def\lsim{\raise0.3ex\hbox{$<$\kern-0.75em\raise-1.1ex\hbox{$\sim$}}}
\def\gsim{\raise0.3ex\hbox{$>$\kern-0.75em\raise-1.1ex\hbox{$\sim$}}}

\def\dAu  {$d$Au}

\def\pp   {$pp$}
\def\pA   {$pA$}

\def\AA   {$AA$}

\def\RCP    {\mbox{$R_{\rm CP}$}}
\def\RdAu    {\mbox{$R_{d\rm Au}$}}

\def\jpsi   {\mbox{$J/\psi$}}
\def\ccbar {\mbox{$c\bar{c}$}}
\def\pT      {\mbox{$p_{T}$}}

\def\sigabs {\mbox{$\sigma_{\mathrm{abs}}$}}
\def\beq     {\begin{equation}}
\def\eeq     {\end{equation}}

\usepackage{ifpdf}
\ifpdf
\usepackage[pdftex]{hyperref}
\else
\usepackage[hypertex]{hyperref}
\fi

\hypersetup{
  pdftitle={},%
  pdfauthor={},%
  pdfsubject={},%
  pdfkeywords={},%
  pdfstartview={},%
  bookmarksopen=true, breaklinks=true, debug=true, %
  colorlinks=true, linkcolor=red, citecolor=blue, urlcolor=blue
}

\journalname{Few Body Systems}
\begin{document}

\title{Centrality, rapidity, and transverse-momentum dependence of gluon shadowing and antishadowing on $J/\psi$ production in $d$Au collisions at $\sqrt{s}$=200 GeV\thanks{Presented at the workshop ``30 years of strong interactions'', Spa, Belgium, 6-8 April 2011.}
}

\author{E.~G.~Ferreiro \and F.~Fleuret \and  J.P.~Lansberg \and N.~Matagne \and A.~Rakotozafindrabe}%

\institute{ E.~G.~Ferreiro \at Departamento de F{\'\i}sica de Part{\'\i}culas, Universidad de Santiago de Compostela, 15782 Santiago de Compostela, Spain 
          \and F.~Fleuret
           \at Laboratoire Leprince Ringuet, \'Ecole Polytechnique, CNRS/IN2P3,  91128 Palaiseau, France 
           \and J.P. Lansberg 
           \at IPNO, Universit\'e Paris-Sud, CNRS/IN2P3, F-91406, Orsay, France
           \and  N.~Matagne \at Universit\'e de Mons, Service de Physique Nucl\'eaire et Subnucl\'eaire, Place du Parc 20, B-7000 Mons, Belgium
           \and A.~Rakotozafindrabe \at IRFU/SPhN, CEA Saclay, 91191 Gif-sur-Yvette Cedex, France
}

\date{Received: date / Accepted: date}

\maketitle

\begin{abstract}
We have carried out a wide study of shadowing and antishadowing effects on \jpsi\ production in
\dAu\ collisions at $\sqrt{s_{NN}}=200$ GeV. We have studied the effects of three different
gluon nPDF sets, using the exact kinematics for a $2\to 2$ process, namely $g+g\to J/\psi+g$
as expected from LO pQCD. We have computed the rapidity dependence of \RCP\ and $R_{d\rm Au}$ for
the different centrality classes of the PHENIX data. For mid rapidities, we have also computed
the transverse-momentum dependence of the nuclear modification factor, which cannot be predicted
with the usual $2\to 1$ simplified kinematics. All these observables have been compared
to the PHENIX data in \dAu\ collisions.
\keywords{$J/\psi$ production \and  heavy-ion collisions \and cold nuclear matter effects \and RHIC}
\end{abstract}

\section{Introduction}
\label{intro}
At high temperature and densities, QCD predicts the existence of a deconfined state of matter, the 
 Quark-Gluon Plasma (QGP) which is expected to be produced in relativistic  nucleus-nucleus 
($AB$) collisions. For thirty years, charmonium production in hadron collisions has been a 
major subject of investigations, on both experimental and theoretical sides. $J/\psi$ production 
should indeed be sensitive to the QGP formation, by a process analogous to Debye screening of 
electromagnetic field in a plasma \cite{Matsui86}. A significant suppression of the $J/\psi$ 
yield was observed at SPS energy by the NA50~experiment~\cite{NA50ref}, and at RHIC by the PHENIX 
experiment in AuAu~\cite{Adare:2006ns} and CuCu~\cite{Adare:2008sh} collisions at 
$\sqrt{s_{NN}}=200\mathrm{~GeV}$. In 2010 and 2011,  data have  been taken at the LHC in PbPb collisions 
at $\sqrt{s_{NN}} = 2.76\mathrm{~TeV}$, where the $J/\psi$ has also been found 
to be suppressed~\cite{Pillot:2011zg,Silvestre:2011ei,Aad:2010px}.

However, the interpretation of the results obtained in $AB$ collisions relies on a good understanding 
and a proper subtraction of the Cold Nuclear Matter~(CNM) effects which are known  to already impact 
the $J/\psi$ production in proton~(deuteron)-nucleus ($pA$ or $dA$) collisions, where the deconfinement 
cannot be reached. Experiments on $d$Au collisions at RHIC \cite{Adare:2007gn,Adare:2010fn} have indeed 
revealed that CNM effects play an essential role at $\sqrt{s_{NN}}=200\mathrm{~GeV}$ in the production
of $J/\psi$ as well as of $\Upsilon$ (see e.g.~\cite{Ferreiro:2011xy}).
In particular, the shadowing of the initial parton distributions due to 
the nuclear environment and the nuclear absorption resulting from the breakup of the $c\bar{c}$ pair by 
its multiple scattering with the remnants of the incident nuclei have a significant impact. 

Previous studies \cite{Ferreiro:2008qj,Ferreiro:2008wc,Ferreiro:2009ur} have also shown that the $J/\psi$ 
partonic-production mechanism affects the way to compute the nuclear shadowing and thus its expected 
impact on the $J/\psi$ production. Most studies on the $J/\psi$ production in hadronic collisions assume 
that the $c\bar{c}$ pair is produced by a \mbox{$2\to 1$} partonic process where both initial particles
 are  two gluons carrying some intrinsic transverse momentum~$k_T$. The sum of the gluon intrinsic $k_T$ 
is transferred to the $c\bar{c}$ pair, thus to the $J/\psi$ since the soft hadronisation process does not 
significantly alter the kinematics. This is supported by the picture of the Colour Evaporation
Model (CEM) at LO (see~\cite{Lansberg:2006dh} and references therein) or of the Colour-Octet (CO) mechanism at
$\alpha_s^2$~\cite{Cho:1995ce}.  In such approaches, the transverse momentum $p_T$ of the
$J/\psi$ comes {\it entirely}  from the intrinsic $k_T$ of the initial gluons. We will refer to this production 
mechanism as to the \emph{intrinsic} scheme.

\begin{figure}[htb!]
\begin{center}
\includegraphics[width=7cm,keepaspectratio]{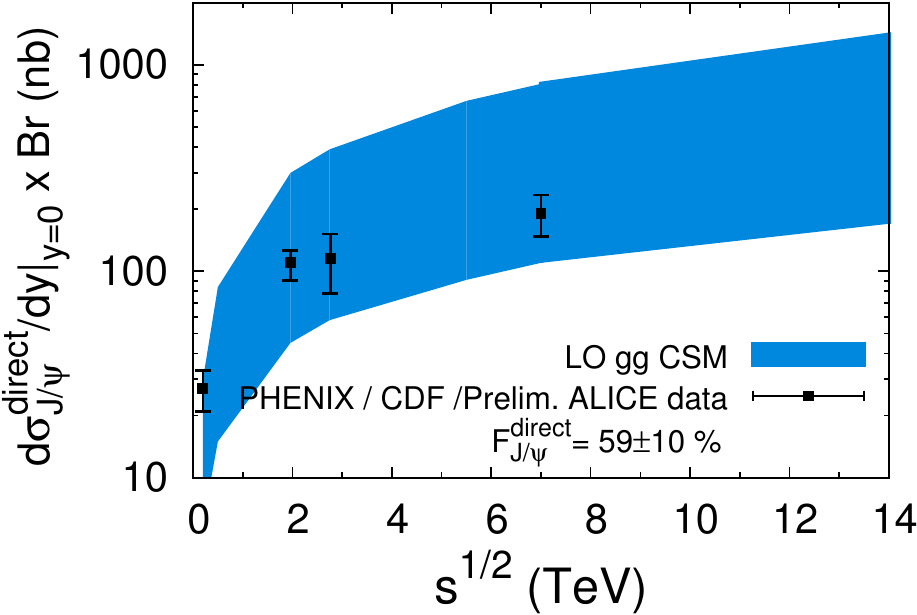}
\caption{$d\sigma^{\mathrm{direct}}_{J/\psi}/dy|_{y=0}\ \times$ Br from $gg$ fusion in $pp$ collisions for 
$\sqrt{s}$ from 200 GeV up to 14 TeV compared to the PHENIX \cite{Adare:2006kf}, CDF \cite{Acosta:2004yw} and ALICE 
data~\cite{Arnaldi:2011er,Aamodt:2011gj}.}
\label{fig:plot-dsigvsdy-y_0-090611} 
\end{center}
\end{figure}

However, this  is not sufficient to describe the $p_T$ spectrum of quarkonia in hadron collisions. Recent theoretical
works incorporating QCD corrections or $s$-channel cut contributions have 
emphasised~\cite{QCD:recentWorks,Brodsky:2009cf,SchannelCutpapers} that the Colour-Singlet (CS) mediated contributions
 are sufficient to describe the experimental data for hadroproduction of both charmonium and bottomonium systems 
without the need of CO contributions. For instance, as 
illustrated by Fig \ref{fig:plot-dsigvsdy-y_0-090611}, the 
yield predicted by the LO CSM \cite{Lansberg:2010cn} reproduces correctly the PHENIX, CDF and ALICE measurements 
without resorting to any colour-octet mechanism nor parameter fitting.
Furthermore, recent works~\cite{ee} focusing on production at $e^+ e^-$ 
colliders have posed stringent constraints on the size of CO contributions, which are the precise ones supporting a 
\mbox{$2\to 1$} hadroproduction mechanism at low $p_T$~\cite{Lansberg:2006dh}. 

As a consequence, $J/\psi$ production at low and mid $p_T$ likely proceeds via a \mbox{$2\to 2$} process, which we refer to as the 
\emph{extrinsic} scheme, such as $g+g \to J/\psi + g$, instead of a \mbox{$2\to 1$} process.  The former 
$2\to 2$ kinematics  is then the most appropriate to derive CNM effects at RHIC, and to provide predictions 
at LHC energy~\cite{Ferreiro:2011rw,Ferreiro:2011fm}.
One could also go further and consider more than two particles in the final state, as expected from the real-emission contributions 
at NLO and NNLO~\cite{QCD:recentWorks}. It is clear from the yield polarisation~\cite{Lansberg:2010vq} that these 
contributions start to dominate for $p_T$ above $1-2 m_c$. The effect of more partons in the final state is to 
increase the difference between the results obtained in both schemes. However the implementation of NLO and NNLO 
codes in a Glauber model with an inhomogeneous shadowing is not yet available.

In this work, we present our results for the 
rapidity and transverse-momentum dependence of the nuclear modification factors, 
\RdAu\ and \RCP\, obtained using the extrinsic scheme for different collision centralities. We compare them 
with the new PHENIX data \cite{Adare:2010fn}.

\section{Our approach}
\label{sec:1}

In order to describe \jpsi\ production in nuclear collisions, our
Monte~Carlo framework~\cite{Ferreiro:2008qj,Ferreiro:2008wc} is based on the probabilistic Glauber model. 
The nucleon-nucleon inelastic cross section at $\sqrt{s_{NN}} = 200\mathrm{~GeV}$ is taken to be 
$\sigma_{NN}=42\mathrm{~mb}$ and the maximum nucleon density to be $\rho_0=0.17\mathrm{~nucleons/fm}^3$. 
We also need to implement the partonic process for the \ccbar\ production model that allows us to 
describe the \pp\ data and the CNM effects.

For a given \jpsi\ momentum (thus for
fixed rapidity~$y$ and \pT), the processes discussed above, \ie\ the intrinsic $g+g \to \ccbar \to J/\psi \,(+X)$
and the extrinsic $g+g \to J/\psi +g$,  will proceed on the average from initial gluons with different Bjorken-$x$. Therefore,
 they will be affected by different shadowing corrections.

In the intrinsic scheme, the measurement of the \jpsi\ momentum in \pp\ collisions completely fixes the longitudinal
 momentum fraction of the initial partons:
\begin{equation}
x_{1,2} = \frac{m_T}{\sqrt{s_{NN}}} \exp{(\pm y)} \equiv x_{1,2}^0(y,p_T),
\label{eq:intr-x1-x2-expr}
\end{equation}
with $m_T=\sqrt{M^2+p_T^2}$, $M$ being the \jpsi\ mass.

In the extrinsic scheme, the knowledge of
the $y$ and \pT\ spectra is enough to fix $x_1$ and $x_2$.
Actually, the presence of a final-state gluon introduces further degrees
of freedom in the kinematics, allowing several $(x_1, x_2)$ for a given set $(y, p_T)$.
 The four-momentum conservation explicitly results in a more complex expression of $x_2$ as a function of~$(x_1,y,p_T)$:
\begin{equation}
x_2 = \frac{ x_1 m_T \sqrt{s_{NN}} e^{-y}-M^2 }
{ \sqrt{s_{NN}} ( \sqrt{s_{NN}}x_1 - m_T e^{y})} \ .
\label{eq:x2-extrinsic}
\end{equation}
Equivalently, a similar expression can be written for $x_1$ as a function of~$(x_2,y,p_T)$.
Models are then {\it mandatory} to compute the proper
weighting of each kinematically allowed $(x_1, x_2)$. This weight is simply
the differential cross section at the partonic level times the gluon PDFs,
\ie\ $g(x_1,\mu_F) g(x_2, \mu_F) \, d\sigma_{gg\to J/\psi + g} /dy \,
dp_T\, dx_1 dx_2 $.
In the present implementation of our code, we are able to use the partonic differential
cross section computed from {\it any} theoretical approach. In this work, we shall use the Colour-Singlet Model (CSM) at LO at LHC energy, which was shown to be compatible~(see Fig. \ref{fig:plot-dsigvsdy-y_0-090611}) \cite{Brodsky:2009cf,Lansberg:2010cn}  
with the magnitude of the \pT-integrated cross-section as given by the PHENIX \pp\ data~\cite{Adare:2006kf}, the CDF $p\bar{p}$ data~\cite{Acosta:2004yw} and the recent LHC \pp\ data at $\sqrt{s_{NN}} = 7\mathrm{~TeV}$~\cite{Aamodt:2011gj}
and $\sqrt{s_{NN}} = 2.76\mathrm{~TeV}$~\cite{Arnaldi:2011er}.

To obtain the yield of $J/\psi$  in \pA\ and \AA\ collisions, a shadowing-correction
factor has to be applied to the \jpsi\ yield obtained from the simple
superposition of the equivalent number of \pp\ collisions.
This shadowing factor can be expressed in terms of the ratios $R_i^A$ of the
nuclear Parton Distribution Functions (nPDF) in a nucleon belonging to a nucleus~$A$ to the
PDF in the free nucleon:
\begin{equation}
\label{eq:shadow-corr-factor}
R^A_i (x,Q^2) = \frac{f^A_i (x,Q^2)}{ A f^{nucleon}_i (x,Q^2)}\ , \ \
i = q, \bar{q}, g \ .
\end{equation}

The numerical parameterisation of $R_i^A(x,Q^2)$ is given for all parton flavours. Since quarkonia 
are essentially produced through gluon fusion at RHIC~\cite{Lansberg:2006dh}, we restrict our 
study to gluon shadowing. Several shadowing parametrisations are available. Here we will 
consider three of them: EKS98 \cite{Eskola:1998df}, EPS08 \cite{Eskola:2008ca} and nDSg at 
LO~\cite{deFlorian:2003qf}. Recently, a new parametrisation with fit uncertainties, 
EPS09~\cite{Eskola:2009uj}, has been made available. Yet, in the case of gluons, the region 
spanned by this parametrisation is approximately bounded by both the nDS and EPS08 values. 
The central curve of EPS09 is also very close to EKS98. We consider sufficient to use only 
EKS98, EPS08 and NDSg. The spatial dependence of the shadowing has been included with a 
shadowing proportional to the local density \cite{Klein:2003dj,vogtprc05}.

The second CNM effect that we take into account concerns the nuclear  absorption. In the framework 
of the probabilistic Glauber model, this effect is usually parametrised by introducing an effective absorption 
cross section~\sigabs. It reflects the break-up of correlated \ccbar~pairs due to inelastic scattering with 
the remaining nucleons from the incident cold nucleus. Here we choose four values of the effective 
absorption cross section ($\sigma_{\mathrm{abs}} = 0, 2.8, 4.2, 6\mathrm{~mb}$) following our
previous works \cite{Ferreiro:2008wc,Ferreiro:2009ur}.

\section{Results}

\subsection{Rapidity and transverse-momentum dependence of \RdAu}

We first present our results for the \emph{nuclear modification factor} \RdAu\ which characterises the $J/\psi$ 
suppression in $d$Au collisions. It is the ratio obtained by normalising the $J/\psi$ yield in $d$Au collisions 
to the $J/\psi$ yield in $pp$ collisions at the same energy times the average number of binary 
inelastic nucleon-nucleon collisions $N_{coll}$:
\begin{equation}
\RdAu= \frac{dN^{J/\psi}_{d\rm Au}}{\langle N_{coll}\rangle dN^{J/\psi}_{pp}}.
\end{equation}

\begin{figure}[htb!]
\begin{center}
\subfloat[EKS98]{\label{rdauEKS98}\includegraphics[width=0.36\textwidth]{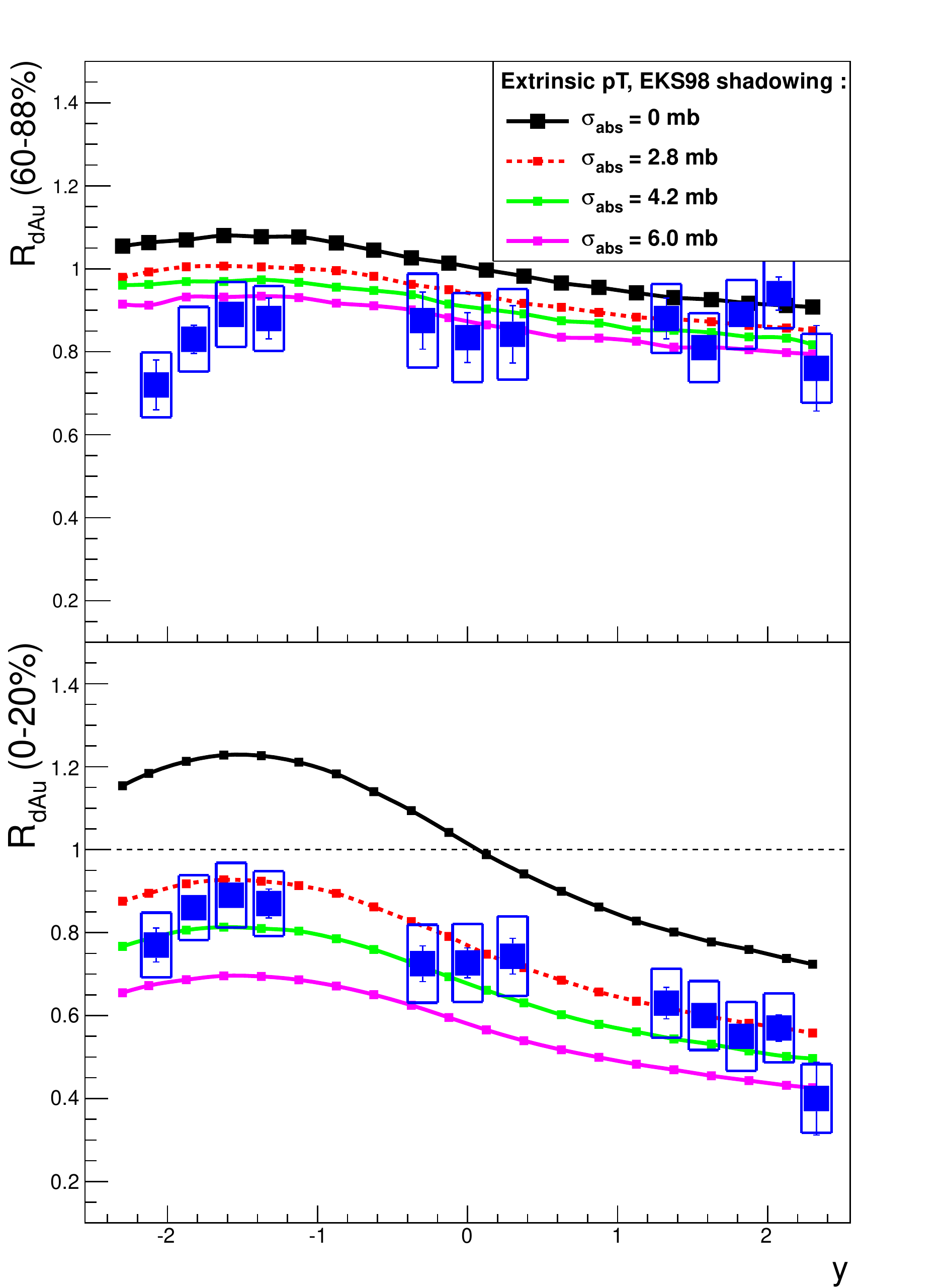}}\hspace*{-0.5cm}
\subfloat[EPS08]{\label{rdauEPS08}\includegraphics[width=0.36\textwidth]{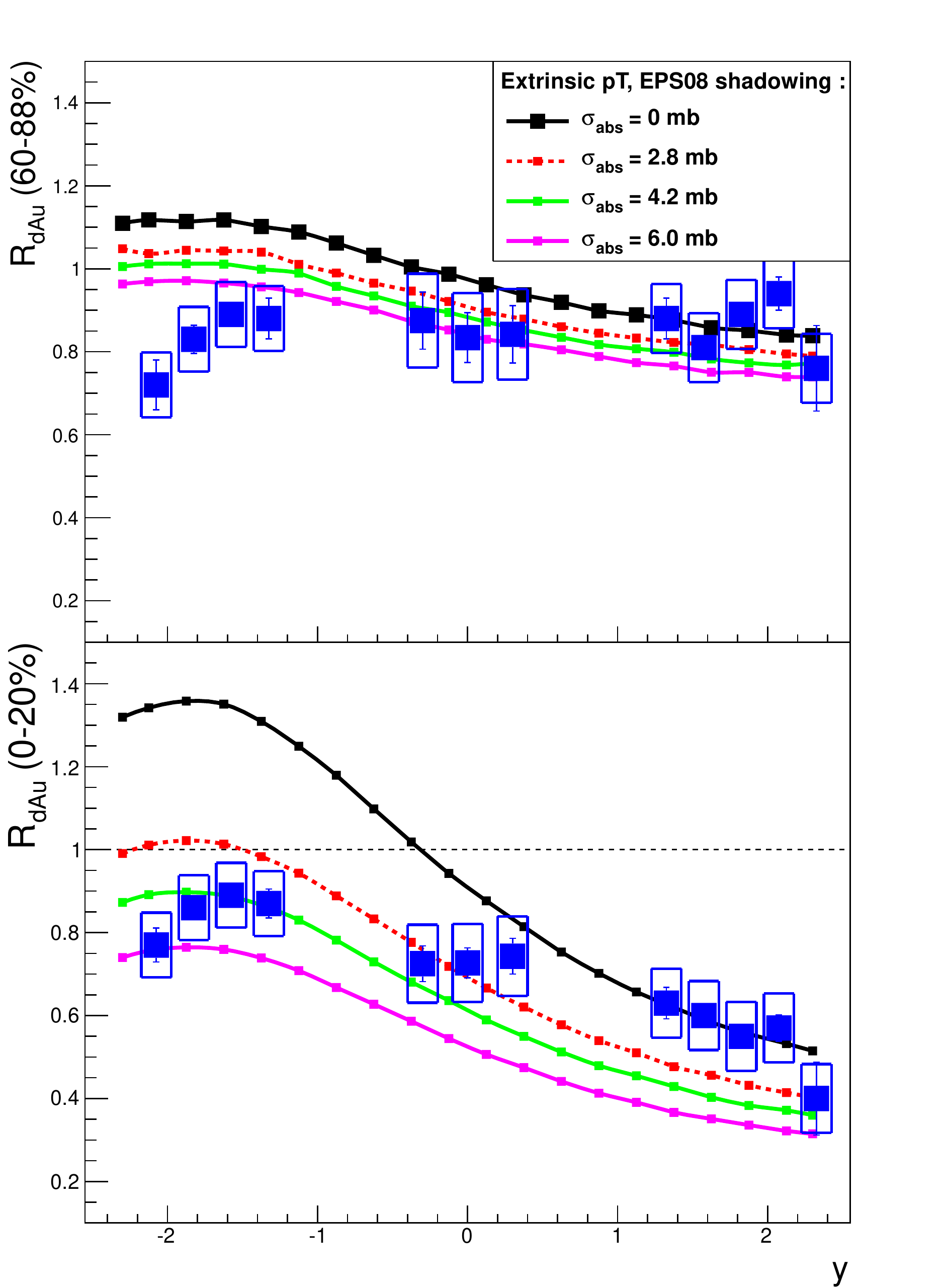}}\hspace*{-0.5cm}
\subfloat[nDSg]{\label{rdauDSG}\includegraphics[width=0.36\textwidth]{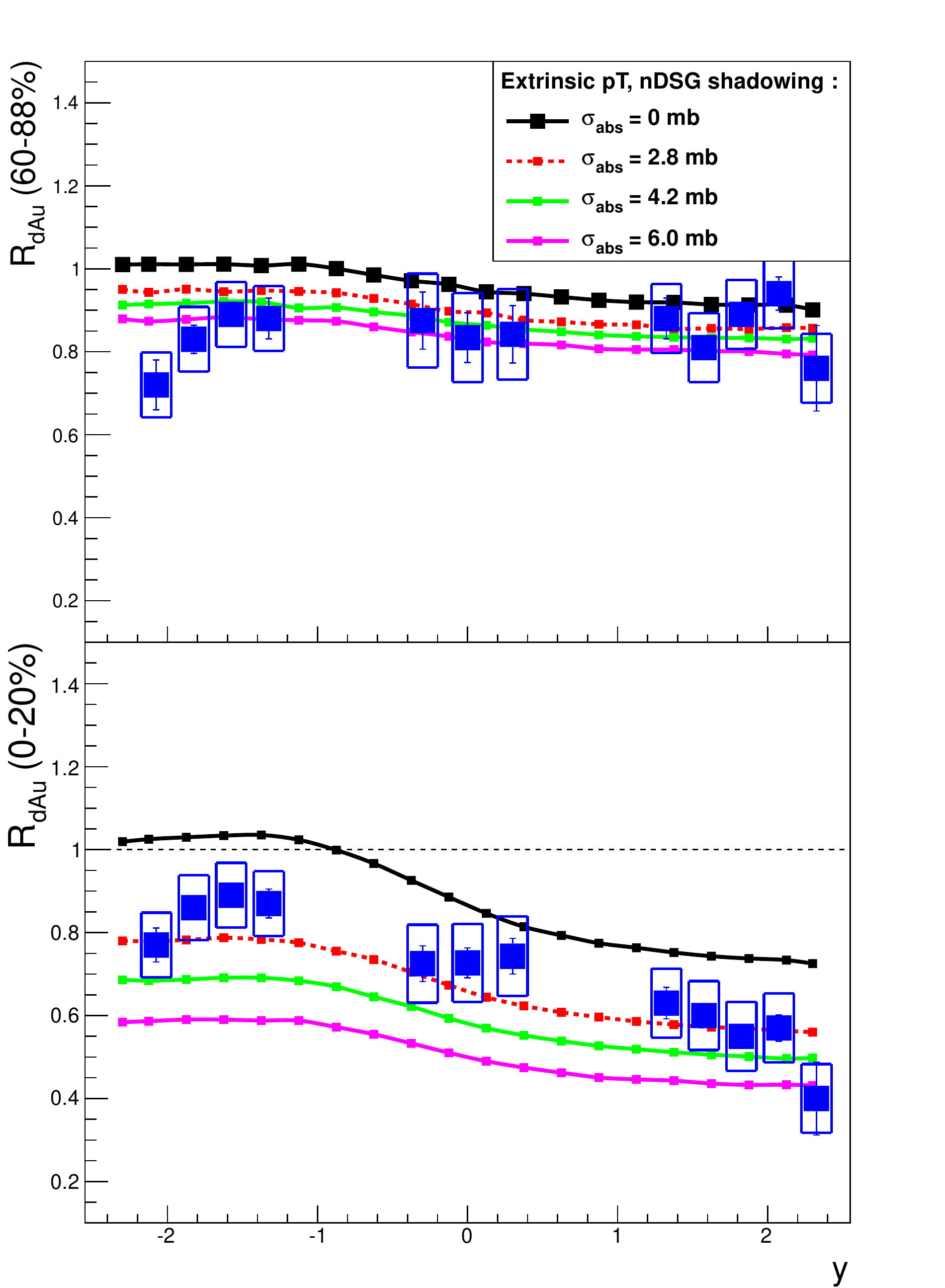}}
\caption{$J/\psi$ nuclear modification factor in $d$Au at $\sqrt{s_{NN}}=200$ GeV for peripheral (upper part) and central (lower part) collisions.
The four curves correspond to different values of the effective absorption cross section using different 
gluon shadowing parametrisations: (a) EKS98, (b) EPS08, (c) nDSg. The data are from PHENIX\protect\footnotemark~\cite{Adare:2010fn}.}
\label{rdau}
\end{center}
\end{figure}

\footnotetext{Note that the PHENIX points showed here do not include  a global systematic uncertainty of $\pm 10\%$ for the peripheral data
and of $\pm 8.5\%$ for the central ones.}

In Fig. \ref{rdau}, we show \RdAu\ vs $y$ obtained for different shadowing parametrisations, EKS98, EPS08 and nDSg. 
We focus only on the extrinsic scheme. Our curves are  compared to the PHENIX data \cite{Adare:2010fn}. 
The lower panels in each of \cf{rdau} (a), (b) and (c) refer to central collisions (centrality class:  0-20 \%, \ie~the 
20 \% most central collisions) and the upper panels to peripheral collisions (centrality class:  60-88 \%).
Our previous study~\cite{Ferreiro:2009ur}, based on older PHENIX data \cite{Adare:2007gn} suggested that the effective absorption 
cross section which reproduced the most accurately the data was $\sigma_{abs} \approx 3-4$ mb.  Among the four different values 
$\sigma_{abs} = 0, 2.8,4.2$ and 6 mb, which we have been considered here, the best match seems to be between 2.8 and 4.2 mb. 
The agreement is good for the most central collisions. For peripheral ones, none of the gluon nPDFs which we used is able to 
accommodate with the most backward data. We also note that the precision of the data does not allow
to distinguish between the different shadowing parametrisations. These results show similar features to those 
Ref. \cite{Adare:2010fn}, where $\sigma_{abs}$  is taken to be 4 mb. This was expected since 
EPS09 shadowing is approximately bounded by EPS08 and nDSg and its central curve is close to EKS98.

We now turn to the discussion of the transverse-momentum dependence of \RdAu\ in the mid-rapidity region. 
Once more, we would like to emphasize that it can only be predicted if one works in the extrinsic scheme. 
Our results for different centrality classes for EKS98 are shown on \cf{rdau_pT_EKS}, for EPS08 on \cf{rdau_pT_EPS} and 
for nDSg on \cf{rdau_pT_DSG}. \RdAu\ is found to increase with $p_T$. This is due to the increase
of $x_2$ for increasing $p_T$ which follows from Eq. \ref{eq:x2-extrinsic}. This effect is more pronounced
 for the EPS08  than for EKS98 and nDSg due to its stronger antishadowing. Note that the centrality
dependence induced by the anti-shadowing -- via its strength dependence on the local nuclear density -- is increasingly
 compensated by that of the break-up probability for increasing $\sigma_{abs}$. For central collisions, 
the production point can be well inside the gold nucleus where the anti-shadowing is expected to be stronger
 but where the break-up probability is also larger.

Our  results are also compared to the most recent PHENIX data \cite{da silva} which suffer from rather large 
experimental uncertainties for increasing $p_T$. The  agreement with the data is reasonable. In addition to the 
plot of \RdAu vs $p_T$ in the mid rapidity region of PHENIX, we show in the appendix our predictions
for backward and forward rapidities to be compared to forthcoming data.

\begin{figure}[htb!]
\begin{center}
\includegraphics[width=5cm,keepaspectratio]{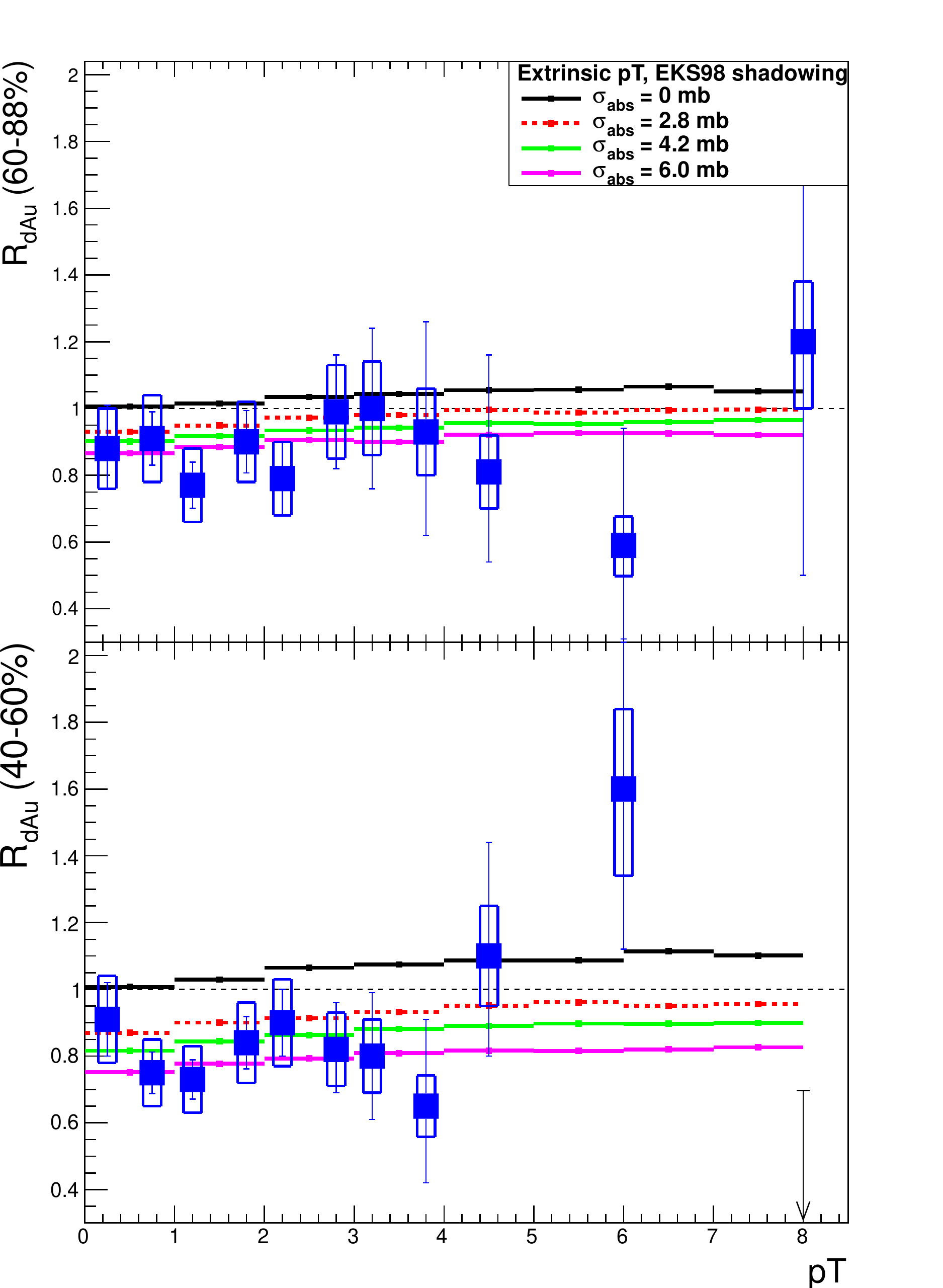}
\includegraphics[width=5cm,keepaspectratio]{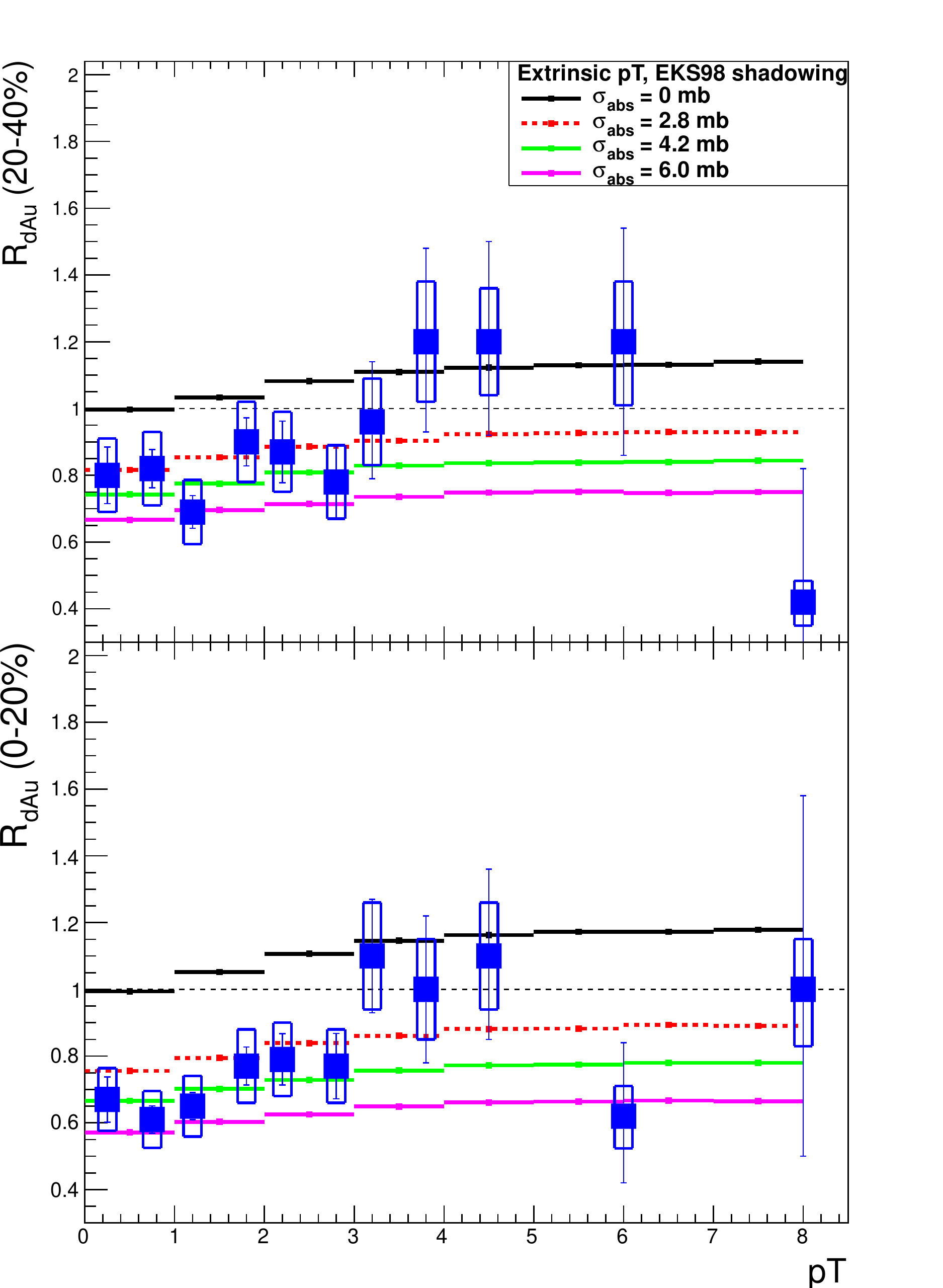}
\caption{$J/\psi$ nuclear modification factor in $dAu$ at $\sqrt{s_{NN}}=200$ GeV vs \pT\ with $|y| < 0.35$ in different 
centrality classes for 
4 effective absorption cross sections using the EKS98 gluon nPDF.}\label{rdau_pT_EKS}
\end{center}
\end{figure}

\begin{figure}[htb!]
\begin{center}
\includegraphics[width=5cm,keepaspectratio]{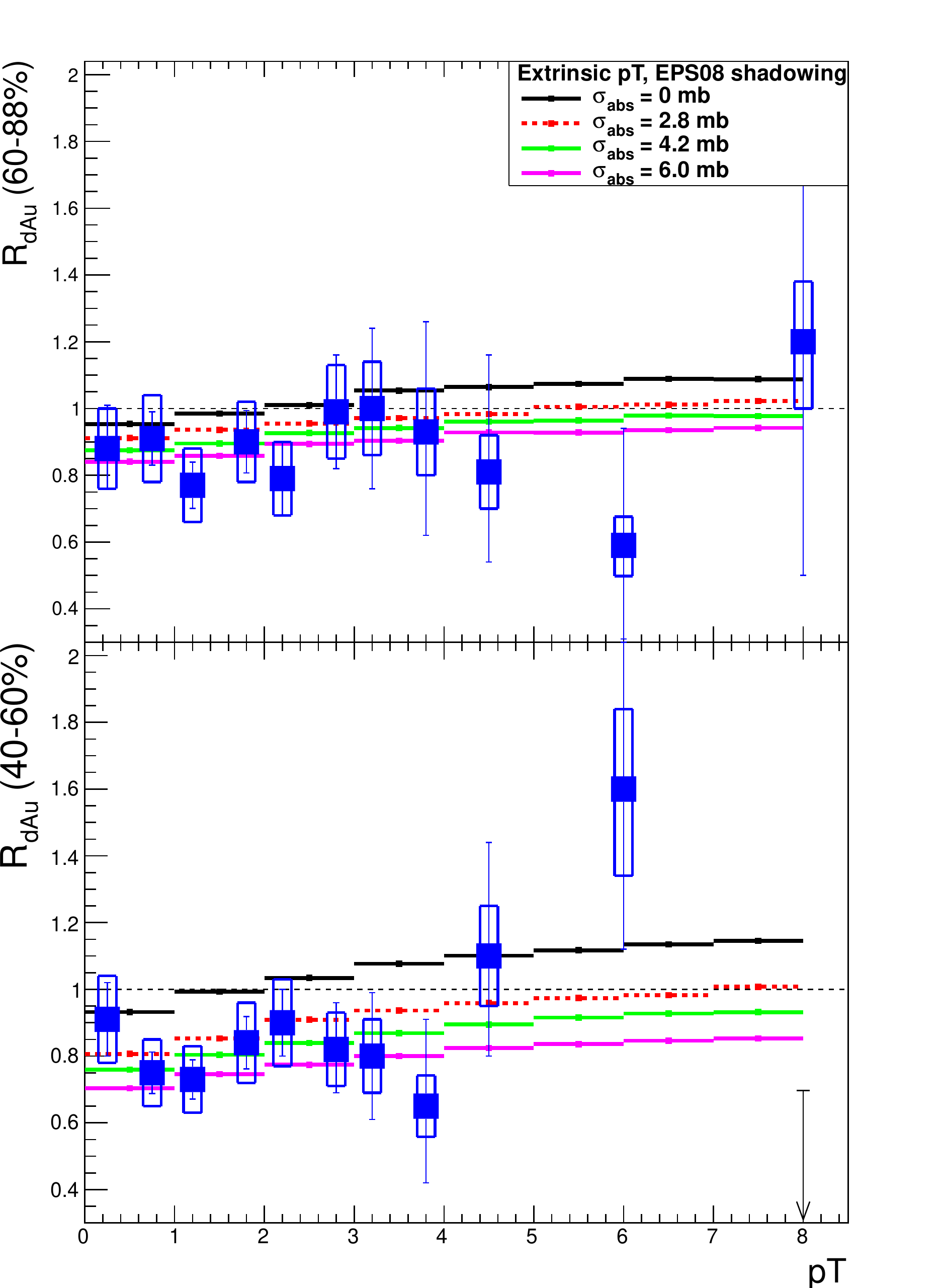}
\includegraphics[width=5cm,keepaspectratio]{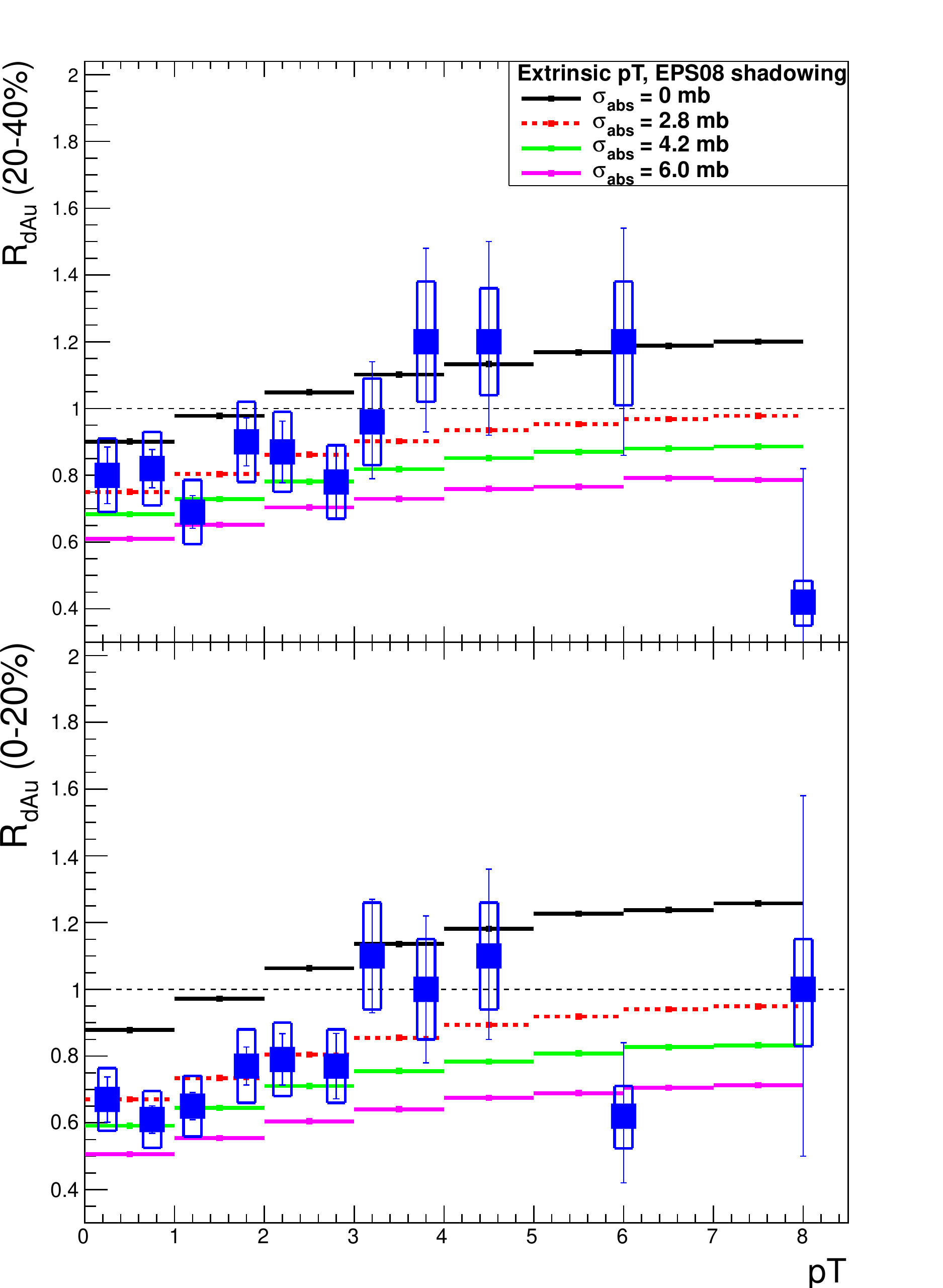}
\caption{Idem as the Fig.~\protect\ref{rdau_pT_EKS} for EPS08.}\label{rdau_pT_EPS}
\end{center}
\end{figure}

\begin{figure}[htb!]
\begin{center}
\includegraphics[width=5cm,keepaspectratio]{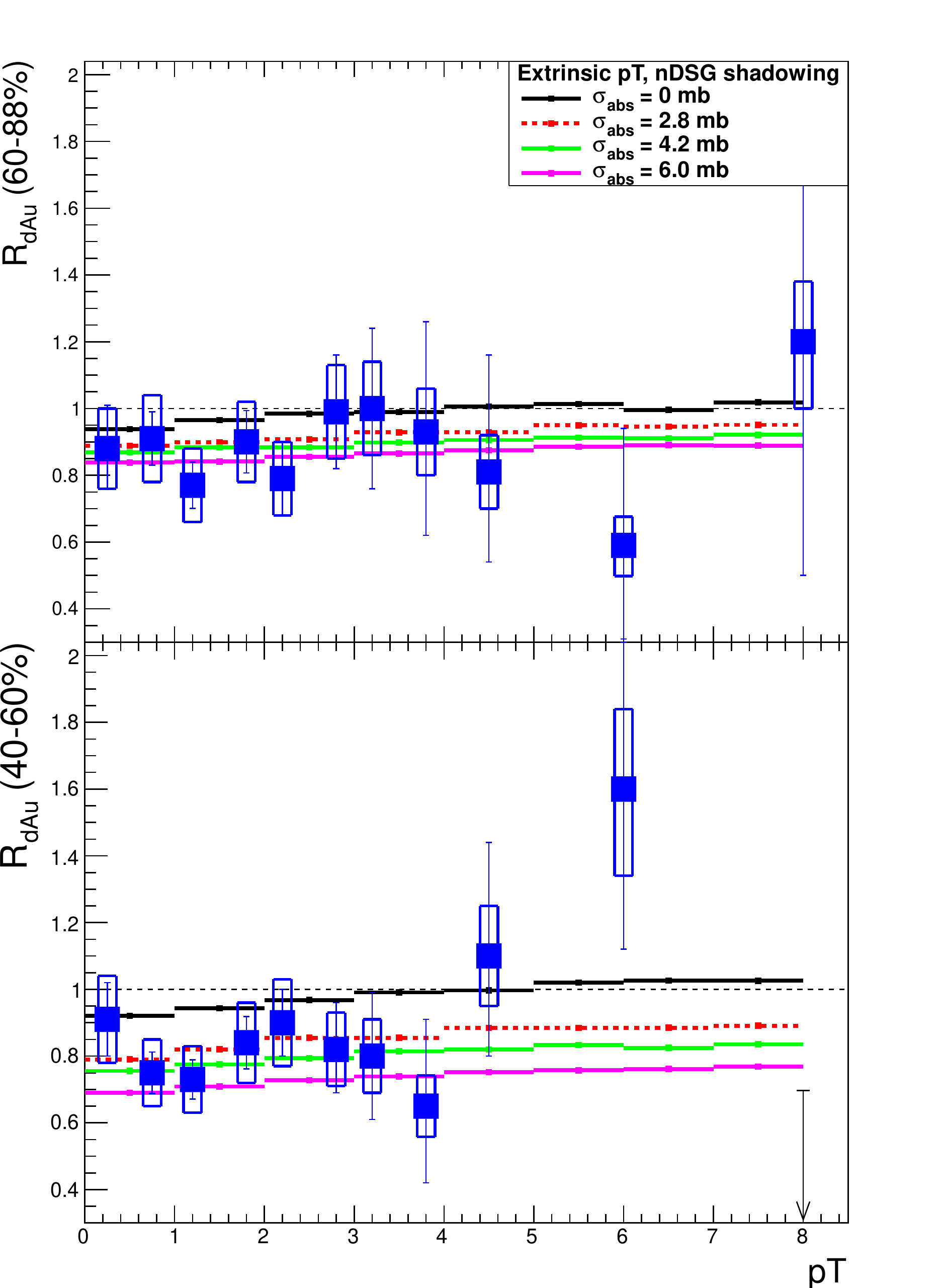}
\includegraphics[width=5cm,keepaspectratio]{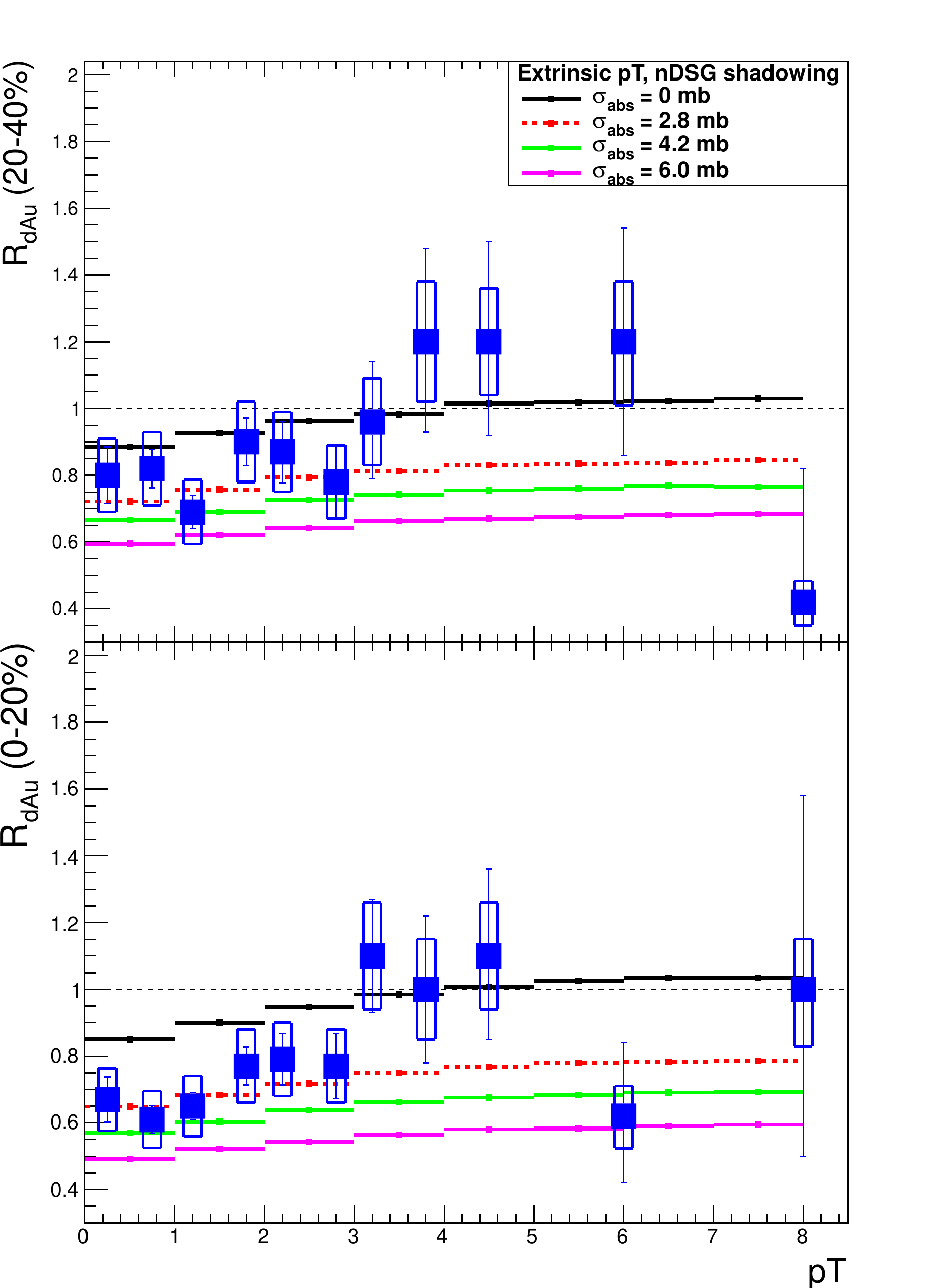}
\caption{Idem as the Fig.~\protect\ref{rdau_pT_EKS} for nDSg.}\label{rdau_pT_DSG}
\end{center}
\end{figure}

\subsection{Rapidity dependence of \RCP}

In this section, we will discuss the rapidity dependence of \RCP\ which give specific information
on the centrality dependence of the CNM. This quantity has the advantage to be a ratio in which 
many of the systematic uncertainties of the data cancel. It is the ratio between central and the peripheral $R_{dAu}$, 
\begin{equation}
 R_{\rm CP} = \left( \frac{\frac{dN^{(0-20\%)}_{J/\psi}}{dy}}{N^{(0-20\%)}_{coll}}\right)\Bigg/\left( \frac{\frac{dN^{(60-88\%)}_{J/\psi}}{dy}}{N^{(60-88\%)}_{coll}}\right)
\end{equation}
\begin{figure}[htb!]
\begin{center} 
\subfloat[EKS98]{\label{rcpEKS98}\includegraphics[width=0.36\textwidth,keepaspectratio]{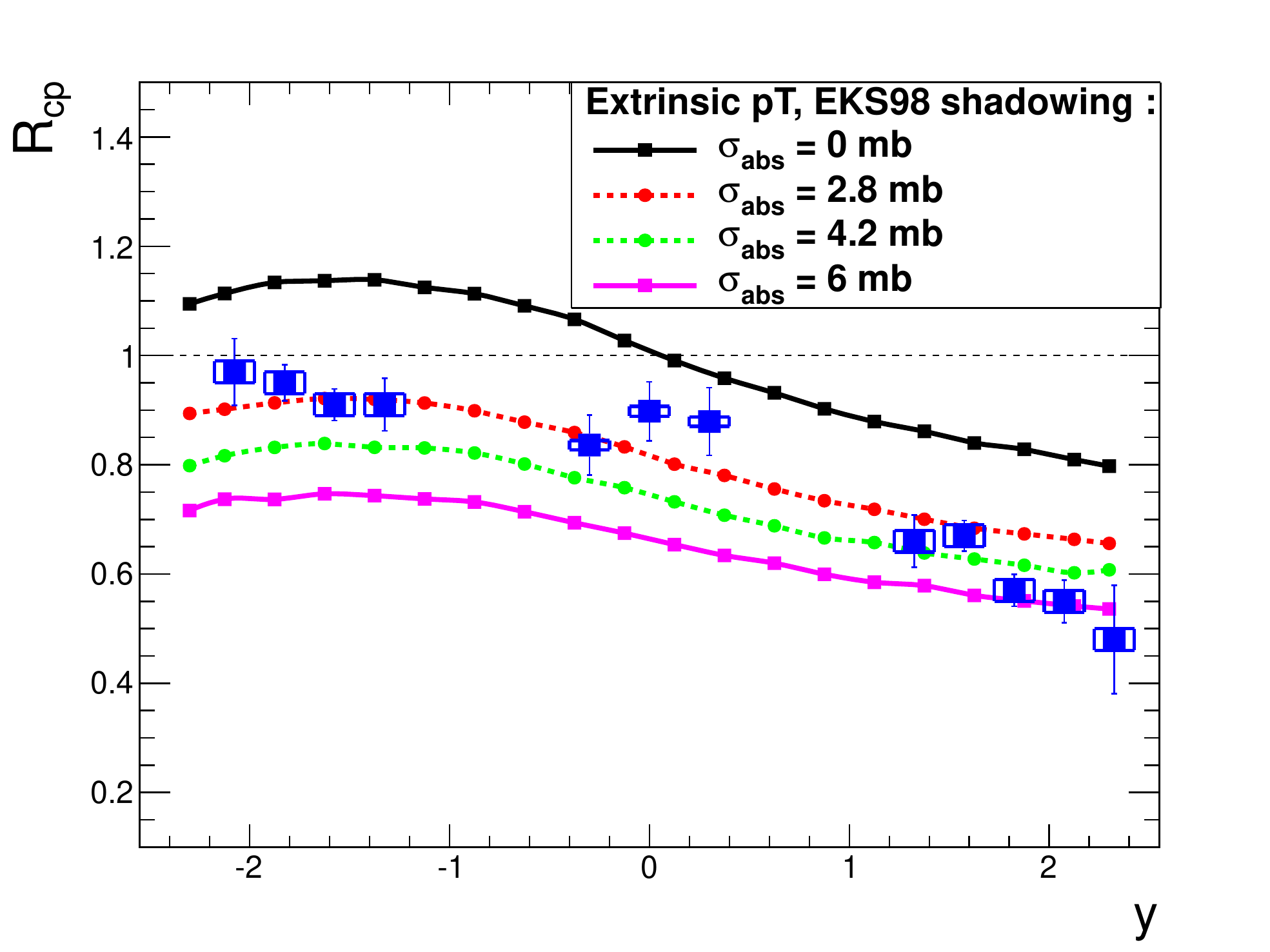}}\hspace*{-0.5cm}
\subfloat[EPS08]{\label{rcpEPS08}\includegraphics[width=0.36\textwidth,keepaspectratio]{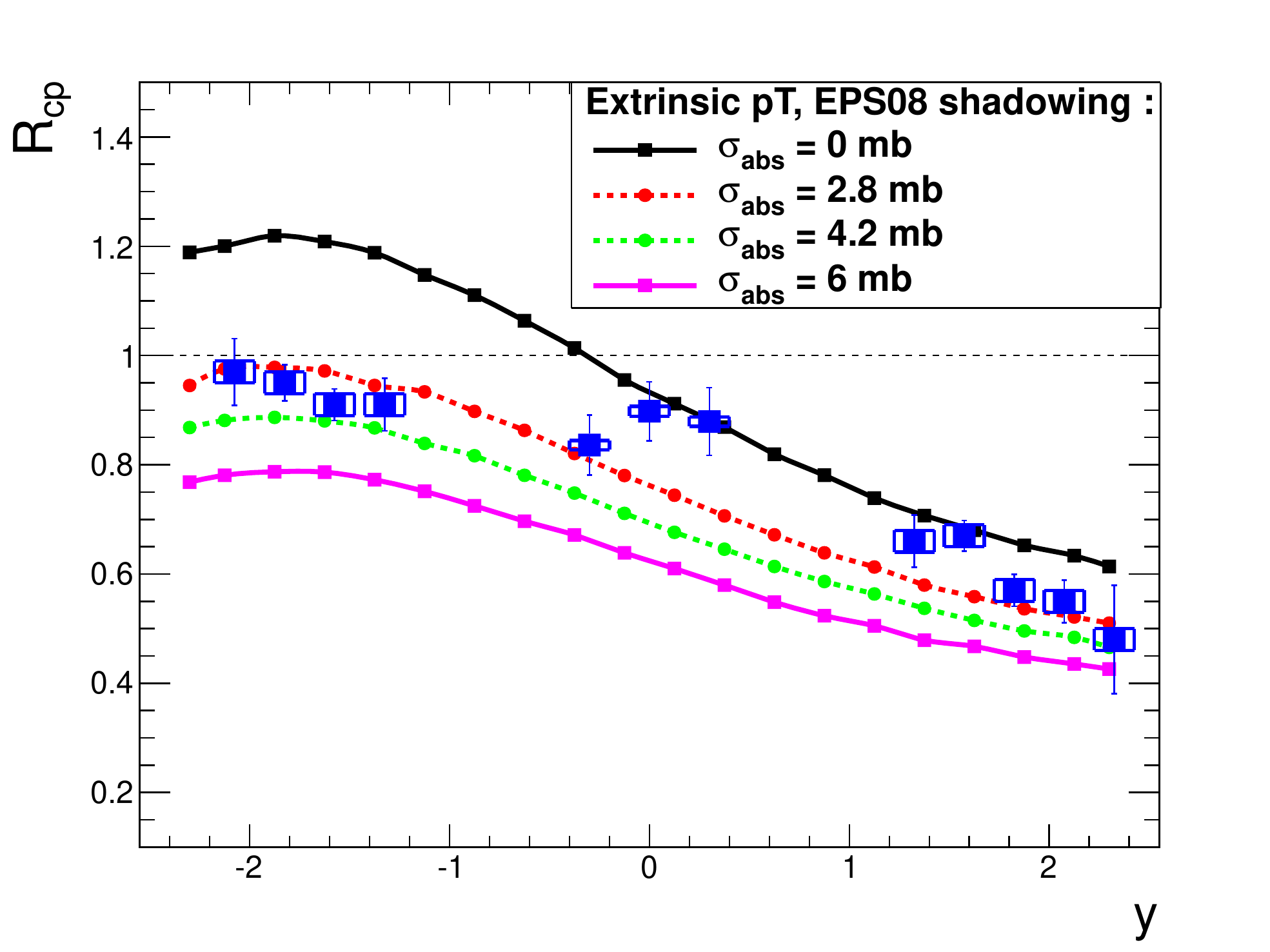}}\hspace*{-0.5cm}
\subfloat[nDSg]{\label{rcpDSG}\includegraphics[width=0.36\textwidth,keepaspectratio]{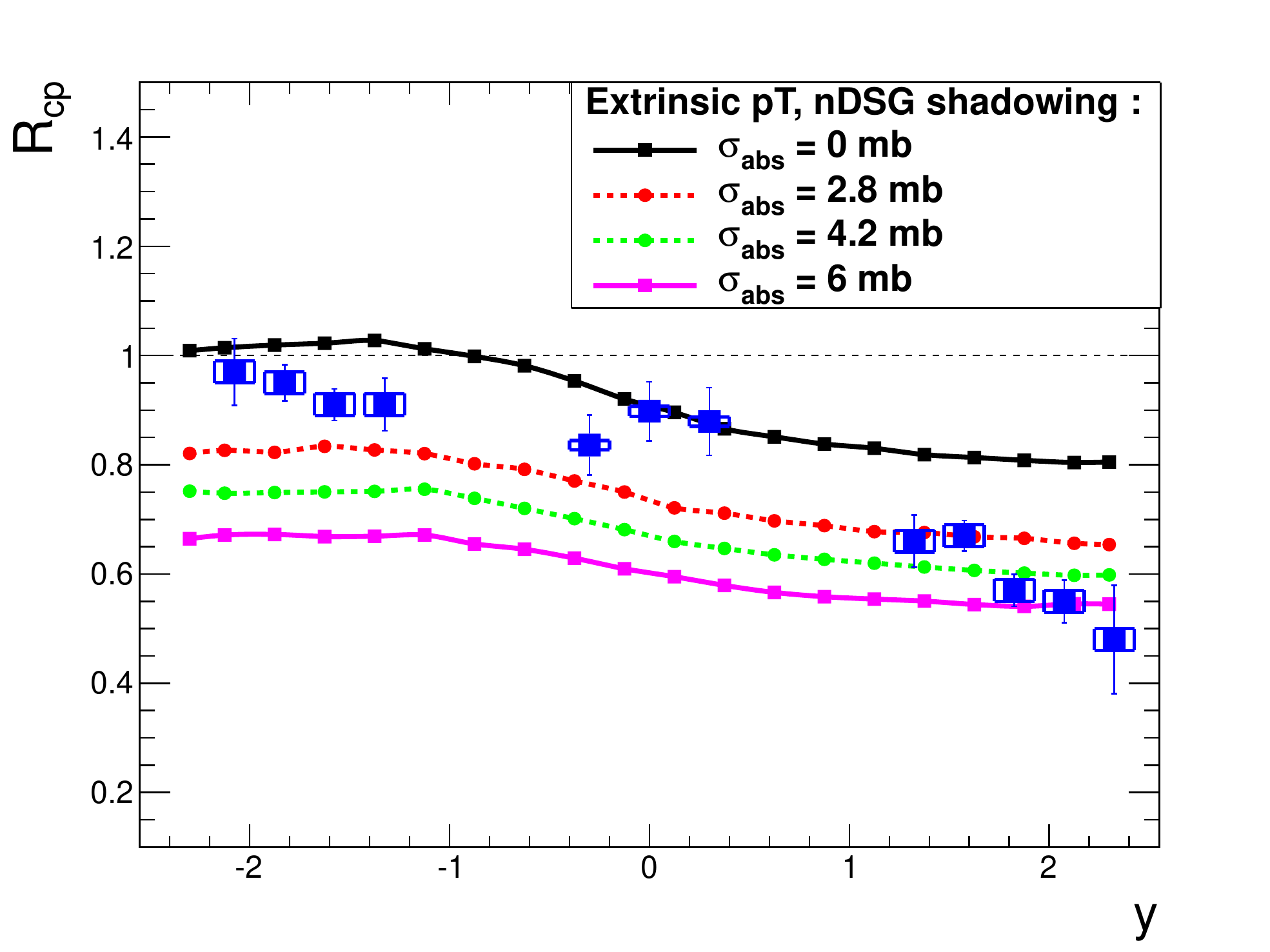}}
\caption{\RCP\ nuclear modification factor in $d$Au collisions at $\sqrt{s_{NN}}= 200$ GeV versus $y$ for 
4 values of $\sigma_{abs}$ using: (a) EKS98, (b) EPS08, (c) nDSg.
The data are from PHENIX\protect\footnotemark~\cite{Adare:2010fn}.}
\label{rcp}
\end{center}
\end{figure}
\footnotetext{Note that the PHENIX points do not include a global systematic uncertainty of $\pm 8.2\%$ .}

Fig. \ref{rcp} presents our results for \RCP\ versus y for three gluon nPDFs (EKS98, EPS08, nDSg) 
and the same four values of $\sigma_{abs}$ as above. We have already discussed the corresponding 
preliminary data from PHENIX in \cite{Ferreiro:2009ur}, from which we performed fits of the 
effective break-up cross section. We had shown at that time that a constant value of $\sigma_{abs}$
was acceptable when the EPS08 nPDF was chosen. As we obtained in our complete fit~\cite{Ferreiro:2009ur} taking into 
account all types of experimental errors~\cite{Ferreiro:2009ur}, 
the comparison with the published PHENIX data shown on Fig. \ref{rcp} 
suggests a $\sigma_{abs}$ smaller than what would be expected from the comparison with $R_{dAu}$ presented 
in the previous section. Our curve for EPS08 seems to better reproduce the most forward points, while it 
slightly misses two of the three mid-$y$ points. A strong shadowing seems in any
case needed to account for these data.

\section{Conclusions}

We have evaluated the rapidity, the centrality and the  transverse-momentum dependence of 
Cold Nuclear Matter effects --essentially the shadowing-- on $J/\psi$ production versus 
rapidity and transverse momentum in $d$Au collisions at $\sigma_{NN}= 200$ GeV and compared our predictions with
the latest PHENIX data. We have used our probabilistic Glauber Monte-Carlo framework, JIN, which allows us
to encode $2\to 2$ partonic mechanisms for $J/\psi$ production. In particular, we have used
the CSM at LO which is now recognised to correctly account for the bulk of the $J/\psi$ cross section in $pp$ at
RHIC.

We have used three gluon nPDFs  (EKS98, EPS08 and nDSg) and considered a reasonable range
of  effective absorption cross sections, $\sigma_{abs}= 0,2.8, 4.2, 6$ mb. Our results, compared to the most recent
 PHENIX data \cite{Adare:2010fn} are in agreement with our previous study \cite{Ferreiro:2009ur} 
where  $\sigma_{abs} \approx 3-4$ mb was suggested from the comparison with $R_{dAu}$ and $\sigma_{abs} \approx 2-3$ mb 
from the comparison with \RCP. This difference may have some physical meaning but the uncertainties both in the
knowledge of gluon (anti-)shadowing and in the experimental data preclude drawing any strong conclusions.
Finally, we reassess that EPS08 with a strength proportional to the local nuclear density 
combined with a $2\to 2$ partonic process is found to reproduce fairly well the most forward \RCP\ data.

\section*{Acknowledgements}
We acknowledge the F.R.S.-FNRS (Belgium), the Ministerio de Ciencia 
(Spain) \& the IN2P3 (France) (AIC-D-2011-0740) and the ReteQuarkonii Networking of the
EU I3 HP 2 program for financial support.  We are very grateful to Darren McGlinchey and Tony
Frawley for providing us with the PHENIX experimental points of \cite{da silva}.


\appendix

\normalsize
\section{Appendix: \RdAu vs $p_T$ for different rapidities and centrality classes}

In addition to the plot of \RdAu vs $p_T$ in the mid rapidity region of PHENIX, we show in this appendix our prediction
for backward and forward rapidities.

\begin{figure}[htb!]
\begin{center} 
\subfloat[EKS98]{\includegraphics[width=0.33\textwidth,keepaspectratio]{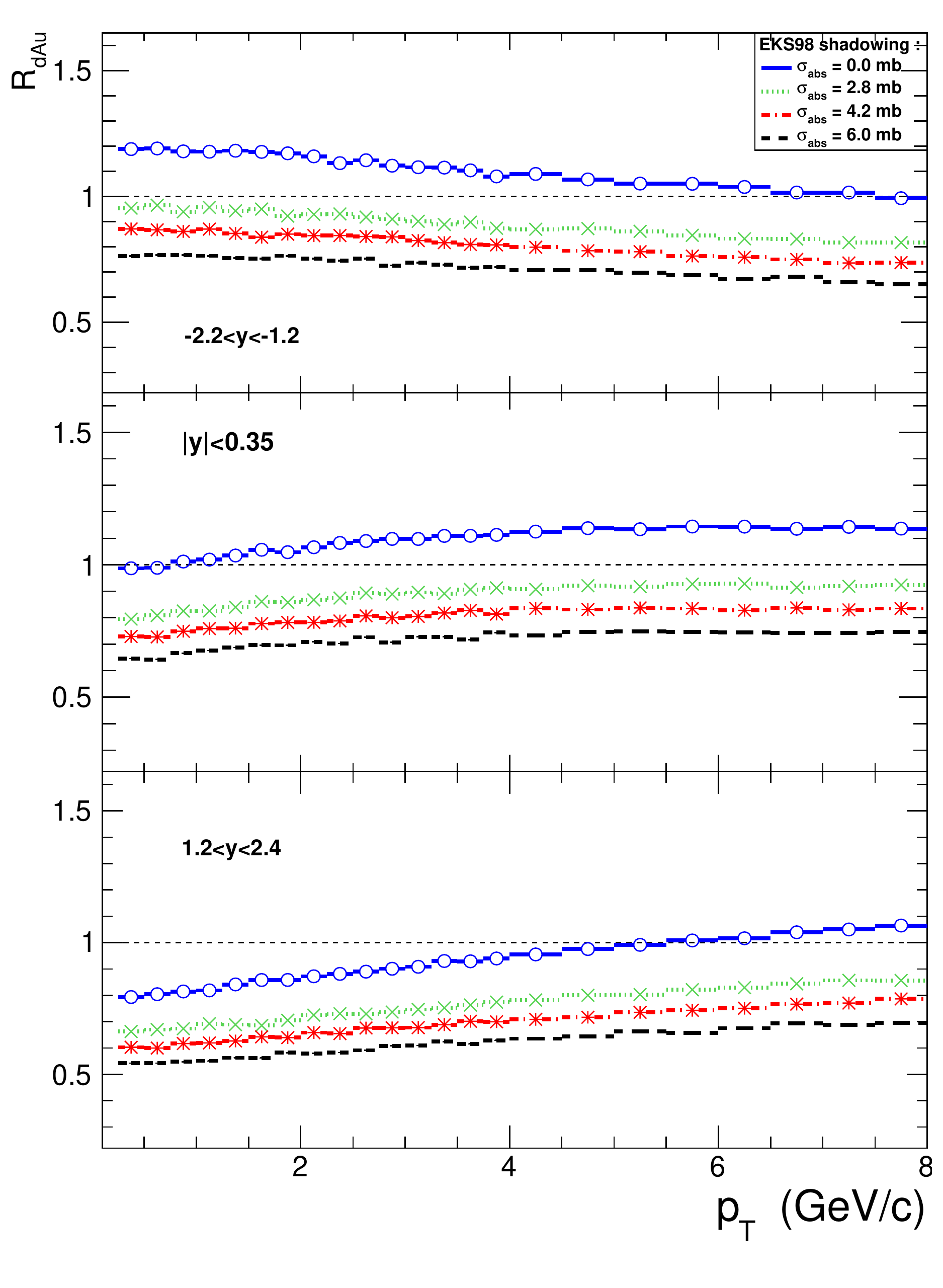}}
\subfloat[EPS08]{\includegraphics[width=0.33\textwidth,keepaspectratio]{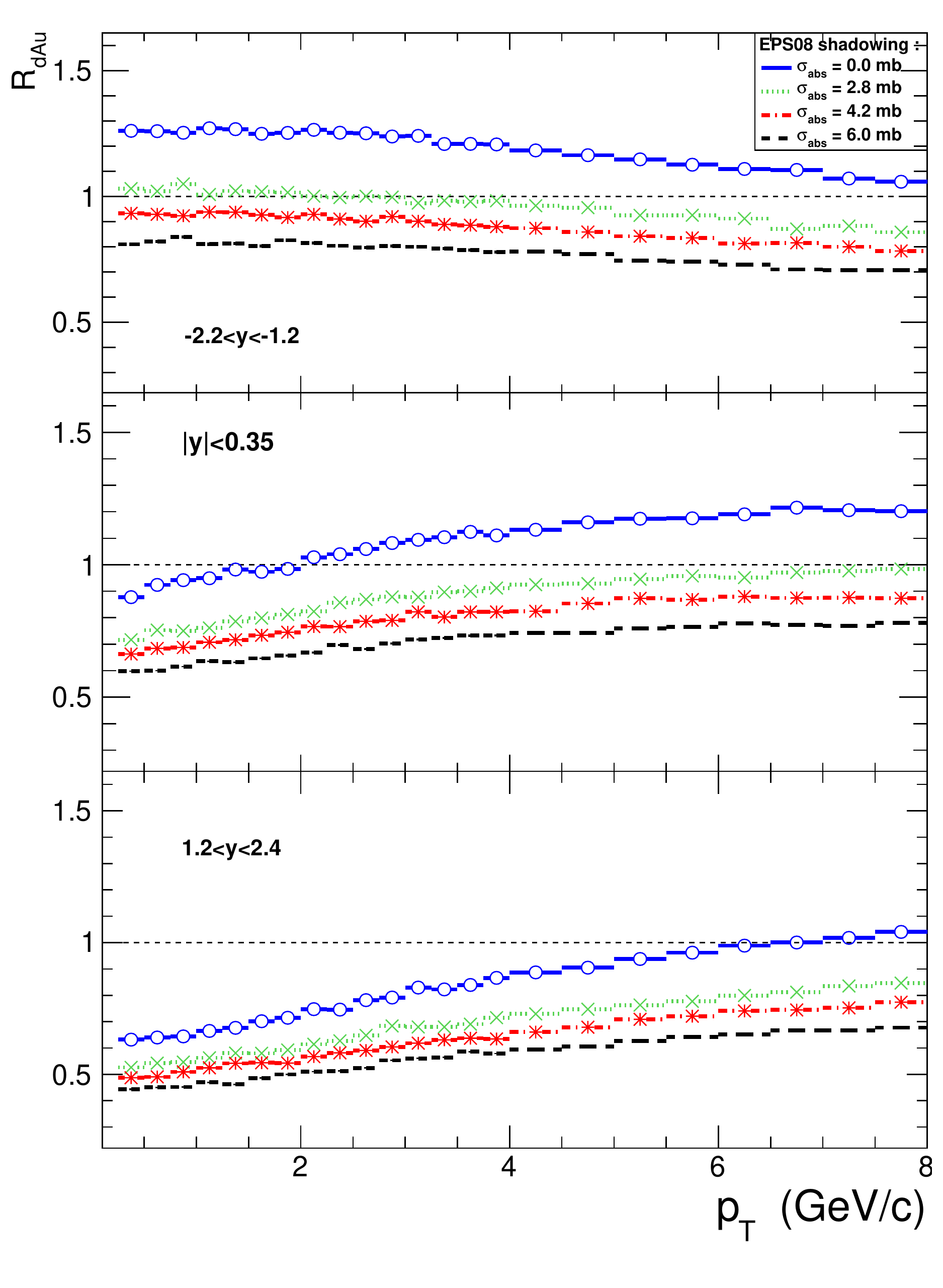}}
\subfloat[nDSg]{\includegraphics[width=0.33\textwidth,keepaspectratio]{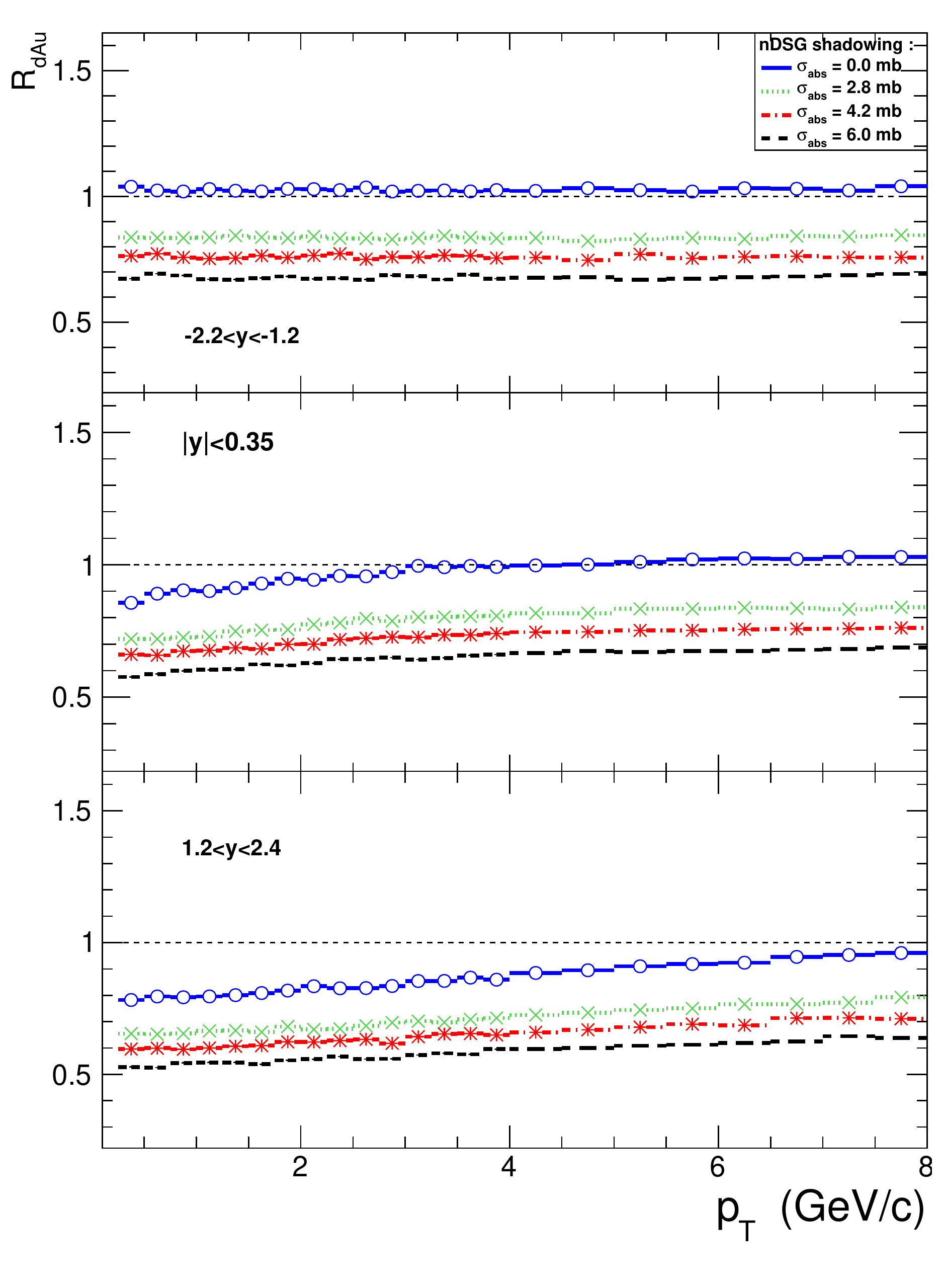}}
\caption{$J/\psi$ nuclear modification factor in $dAu$ at $\sqrt{s_{NN}}=200$ GeV vs \pT\ integrated on the centrality, 
for four effective absorption cross sections  using a) EKS98, b) EPS08, c) nDSg in the 3 rapidity regions covered by
PHENIX}
\label{R_dAu_flatpT_3_y_ranges}
\end{center}
\end{figure}

\begin{figure}[htb!]
\begin{center} 
\subfloat[EKS98]{\includegraphics[width=0.33\textwidth,keepaspectratio]{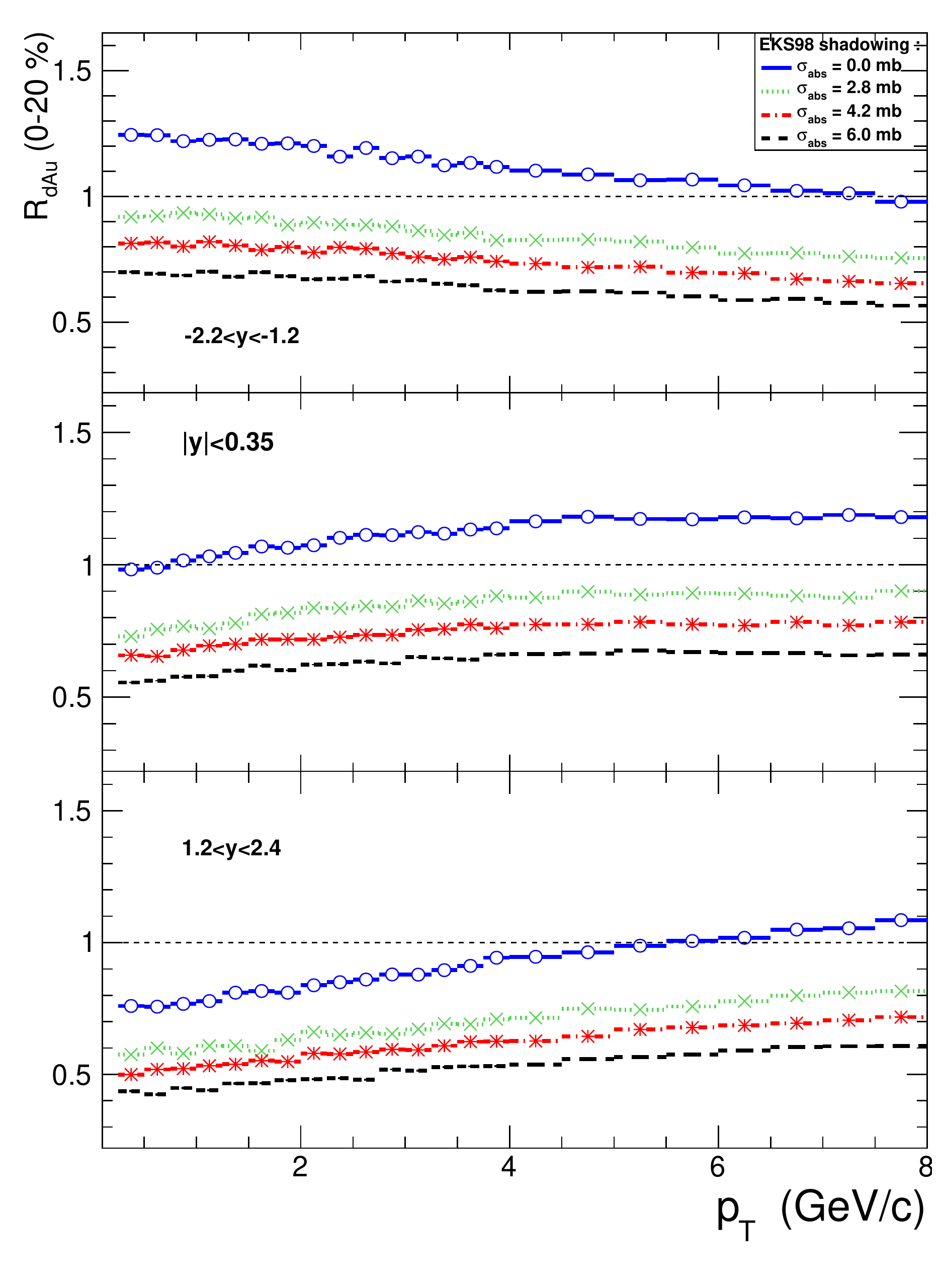}}
\subfloat[EPS08]{\includegraphics[width=0.33\textwidth,keepaspectratio]{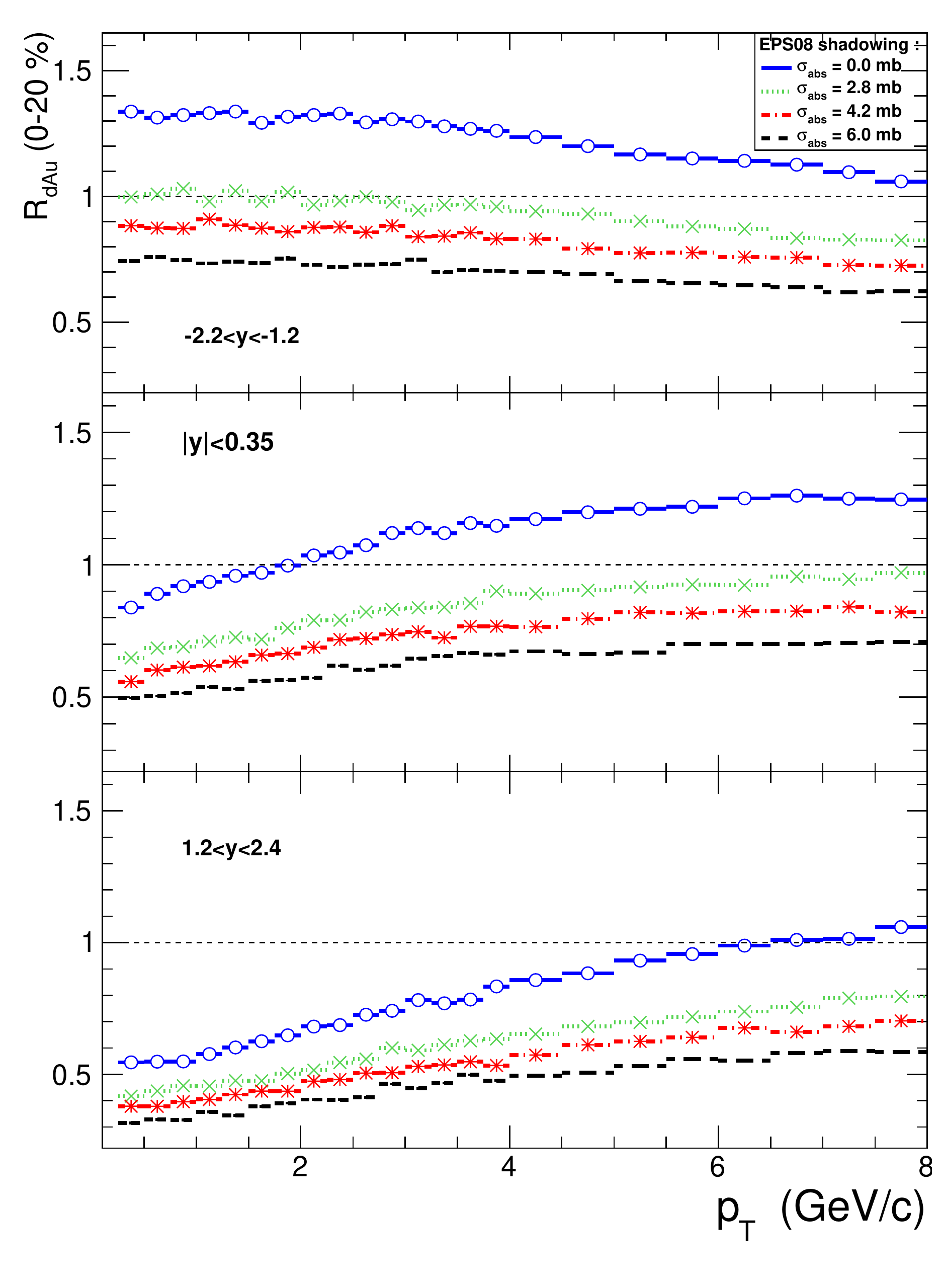}}
\subfloat[nDSg]{\includegraphics[width=0.33\textwidth,keepaspectratio]{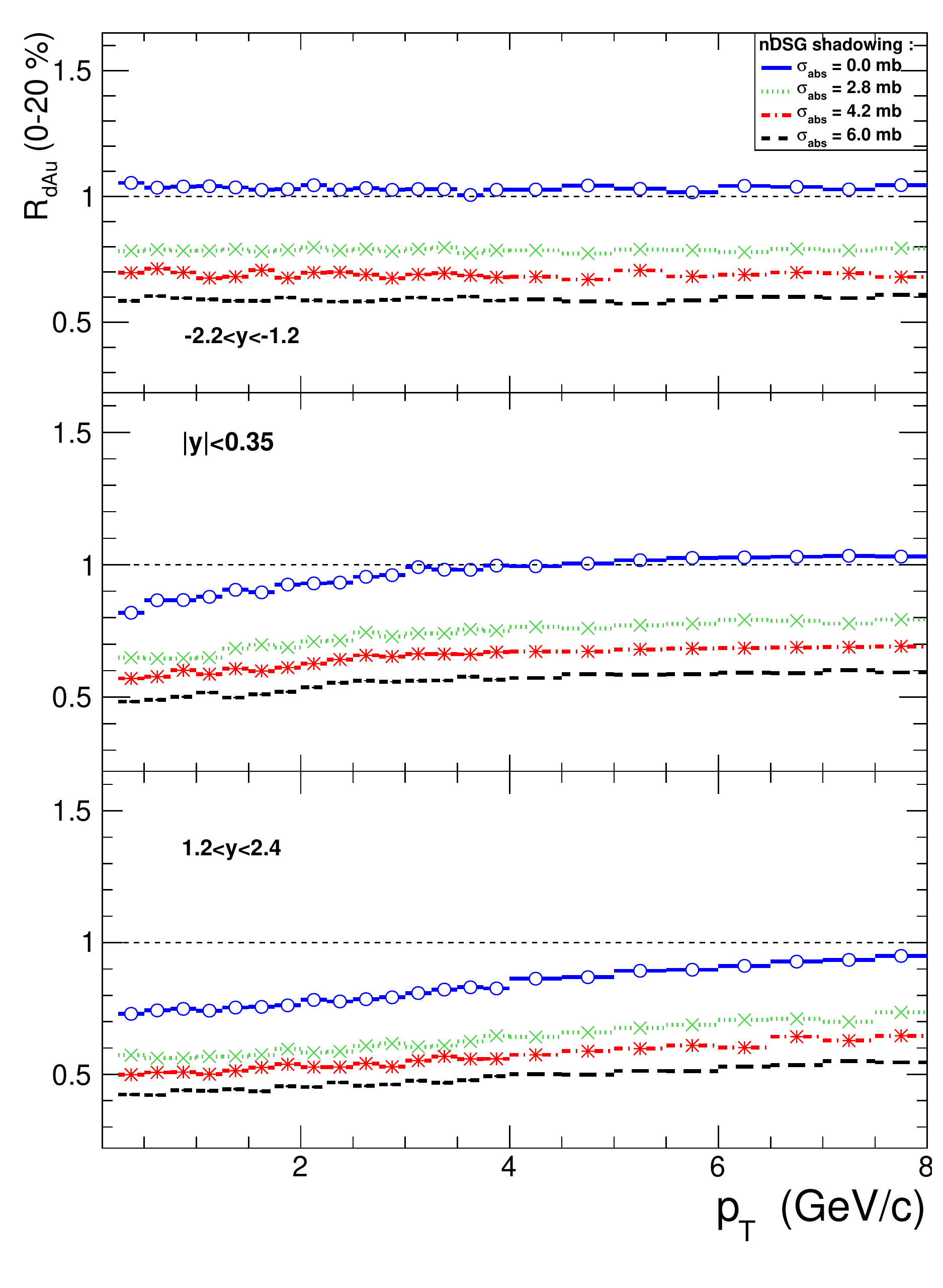}}
\caption{Idem as the Fig.~\protect\ref{R_dAu_flatpT_3_y_ranges} for the centrality class 0-20 \%.}
\end{center}
\end{figure}

\begin{figure}[htb!]
\begin{center} 
\subfloat[EKS98]{\includegraphics[width=0.33\textwidth,keepaspectratio]{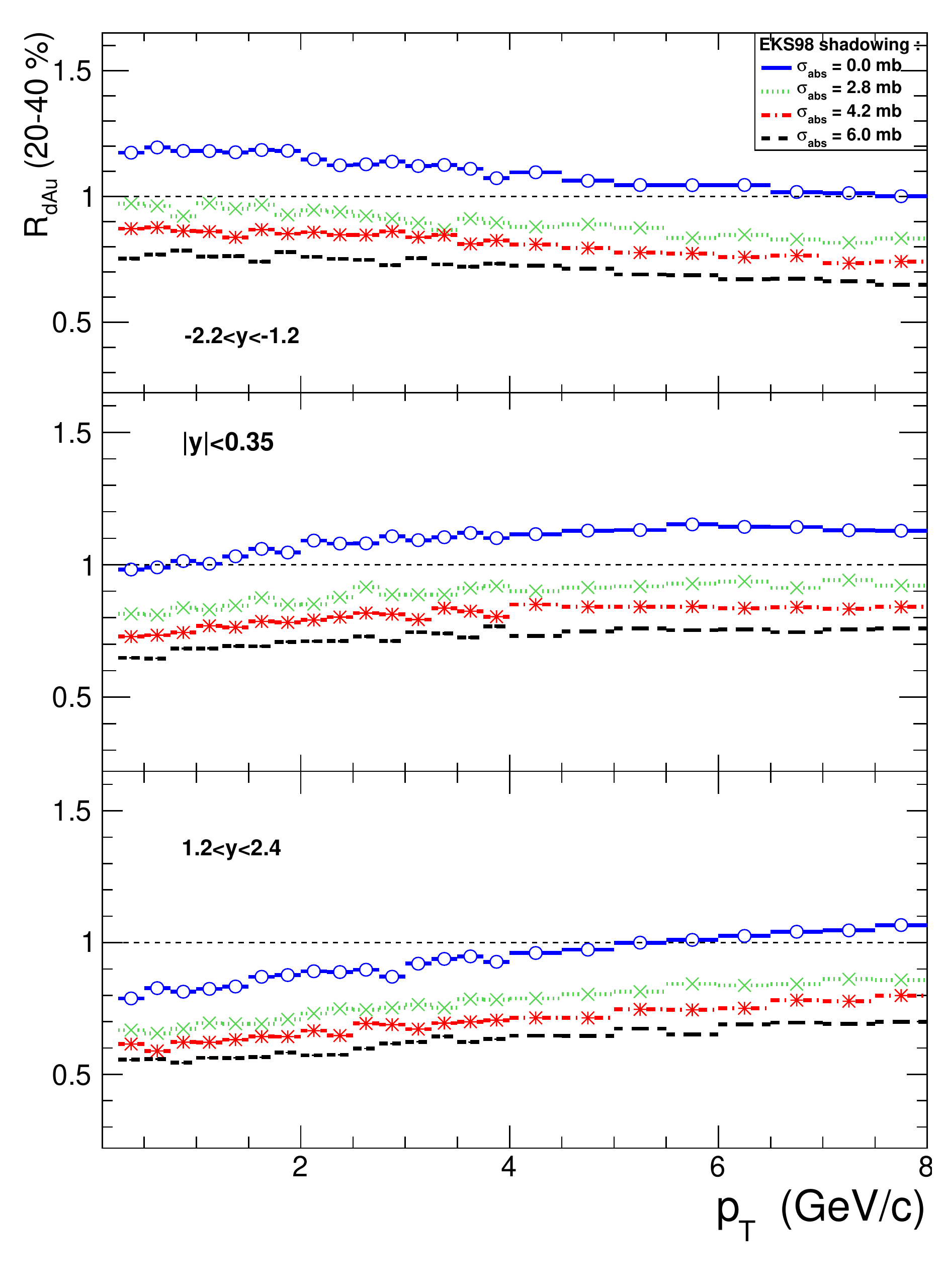}}
\subfloat[EPS08]{\includegraphics[width=0.33\textwidth,keepaspectratio]{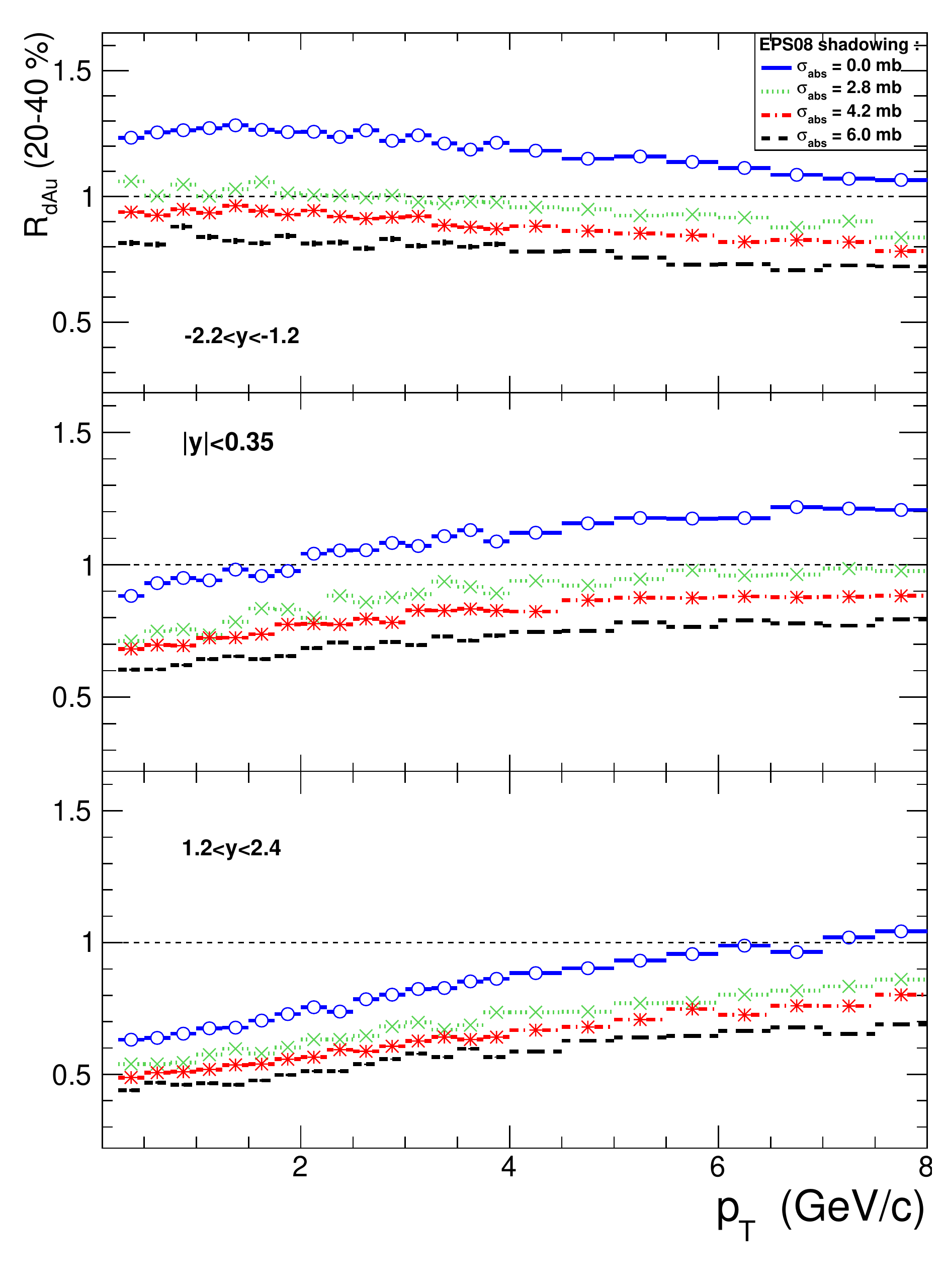}}
\subfloat[nDSg]{\includegraphics[width=0.33\textwidth,keepaspectratio]{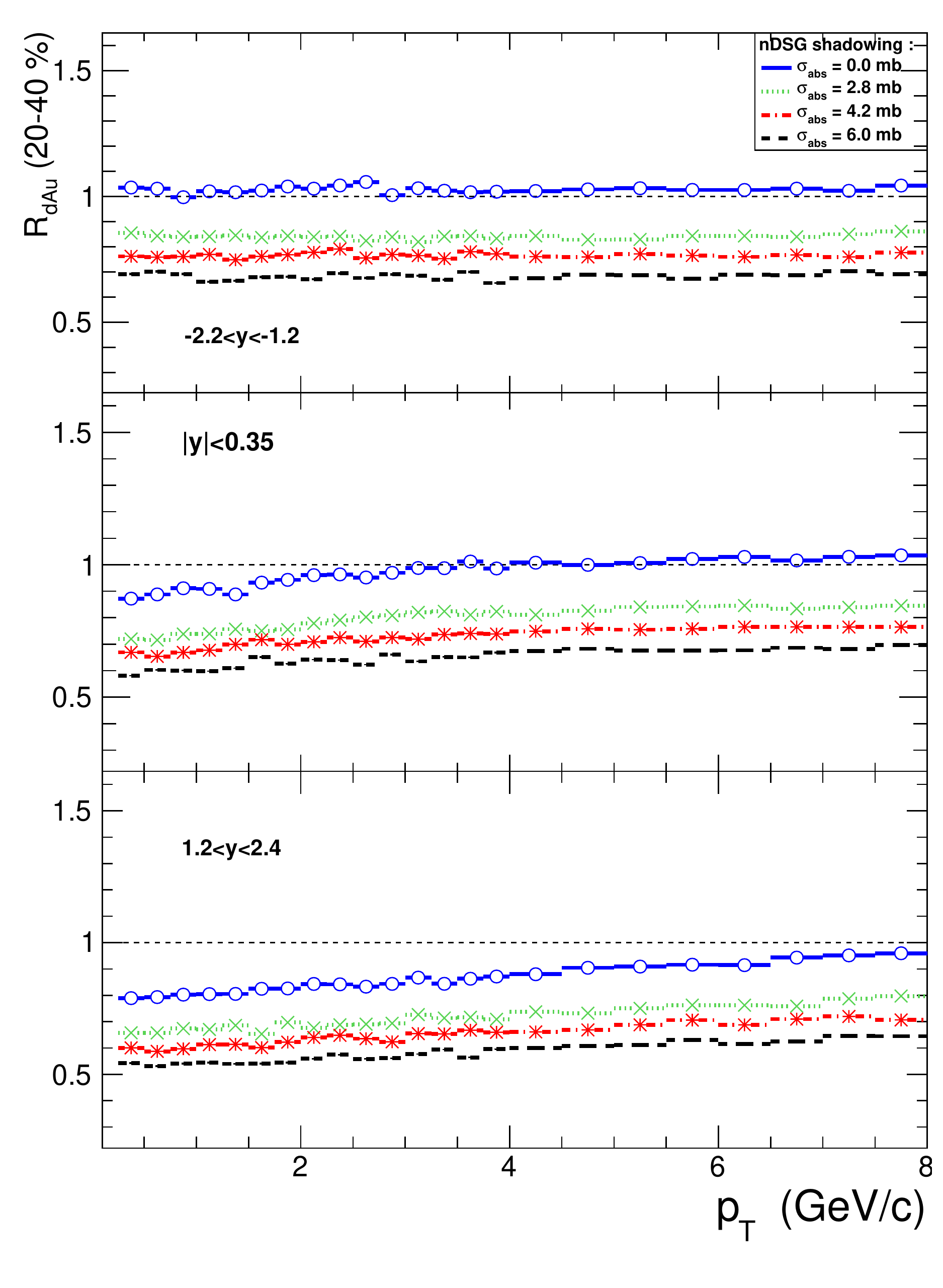}}
\caption{Idem as the Fig.~\protect\ref{R_dAu_flatpT_3_y_ranges} for the centrality class 20-40 \%.}
\end{center}
\end{figure}

\begin{figure}[htb!]
\begin{center} 
\subfloat[EKS98]{\includegraphics[width=0.33\textwidth,keepaspectratio]{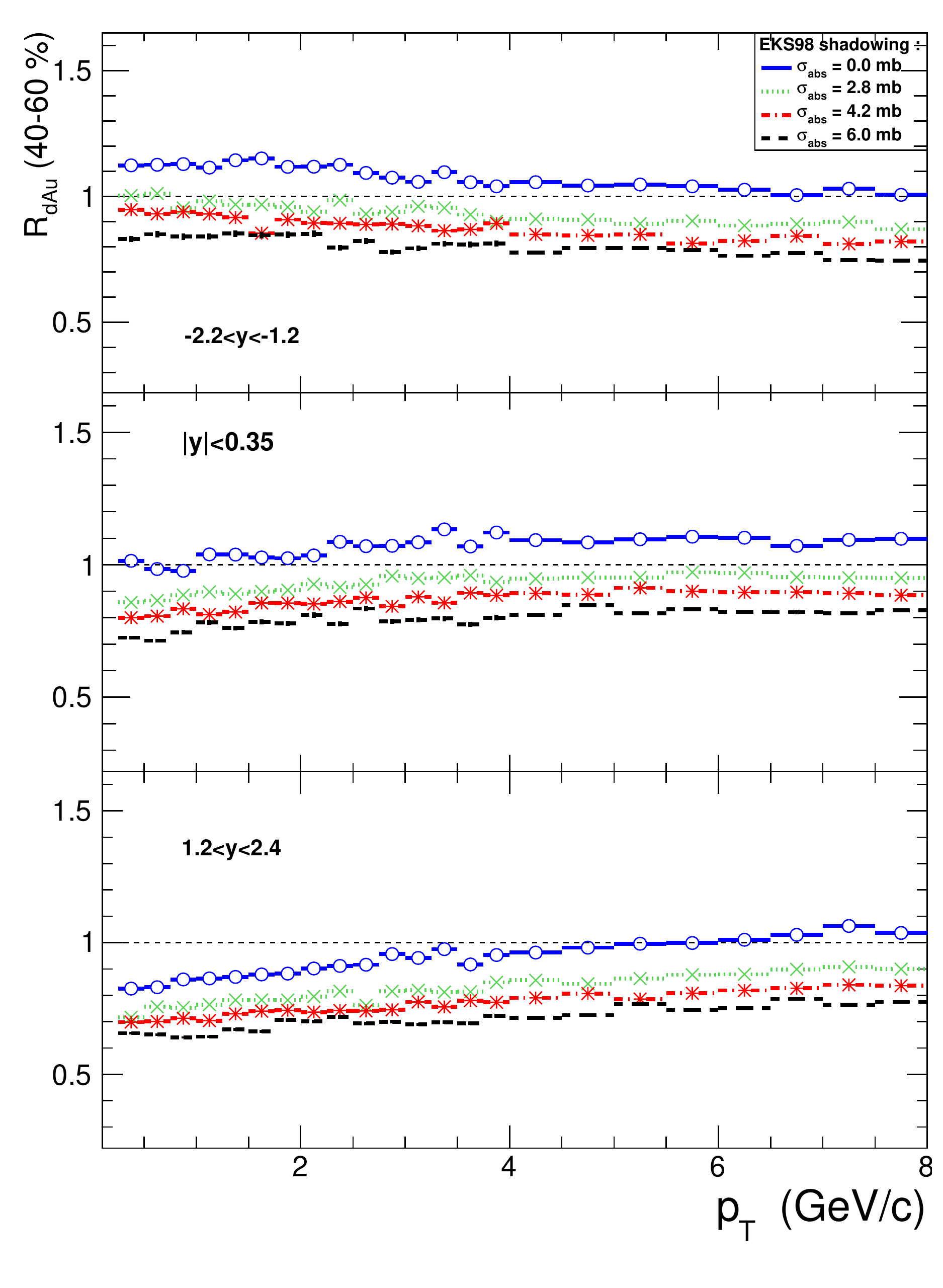}}
\subfloat[EPS08]{\includegraphics[width=0.33\textwidth,keepaspectratio]{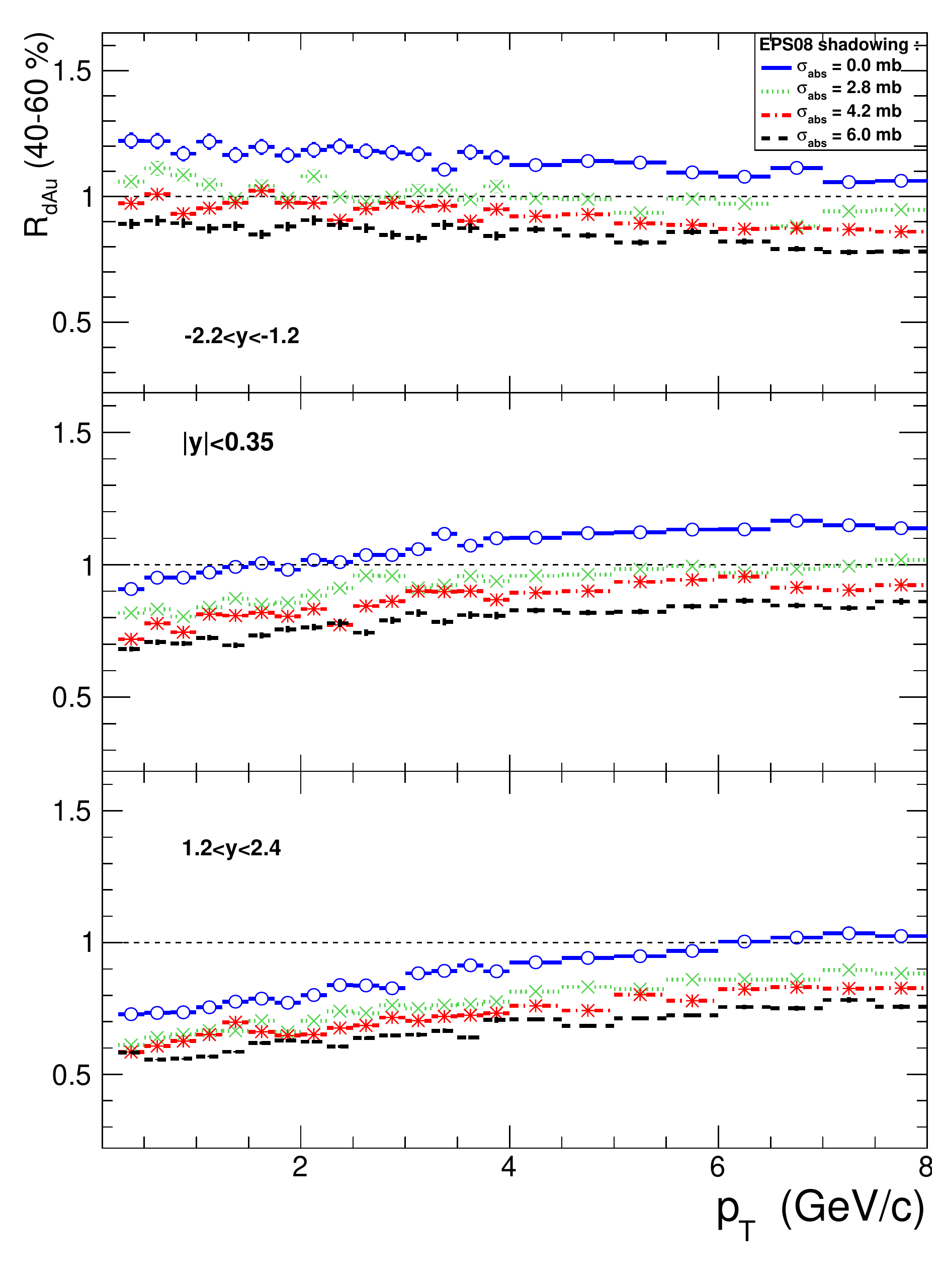}}
\subfloat[nDSg]{\includegraphics[width=0.33\textwidth,keepaspectratio]{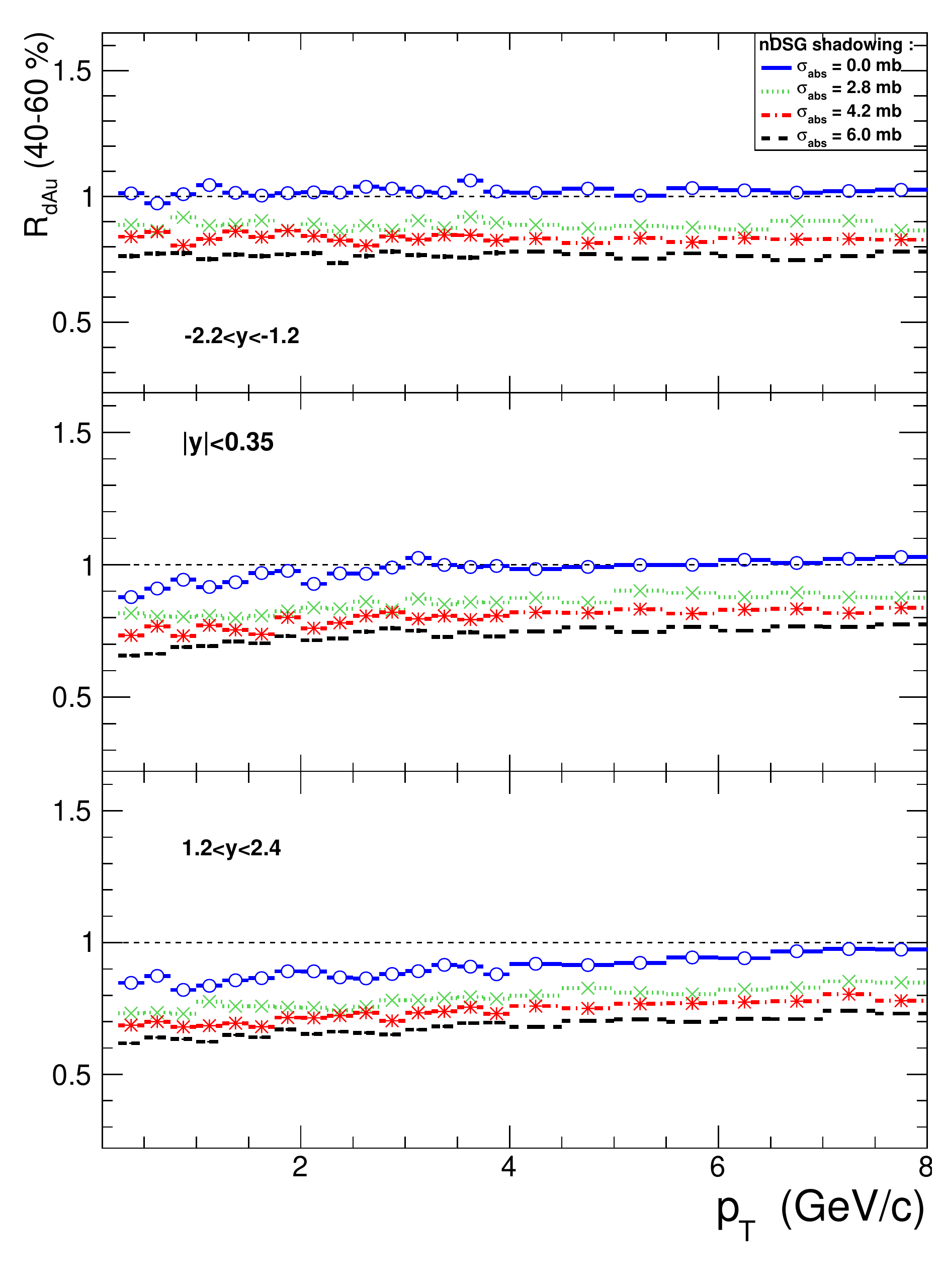}}
\caption{Idem as the Fig.~\protect\ref{R_dAu_flatpT_3_y_ranges} for the centrality class 40-60 \%.}
\end{center}
\end{figure}

\begin{figure}[htb!]
\begin{center} 
\subfloat[EKS98]{\includegraphics[width=0.33\textwidth,keepaspectratio]{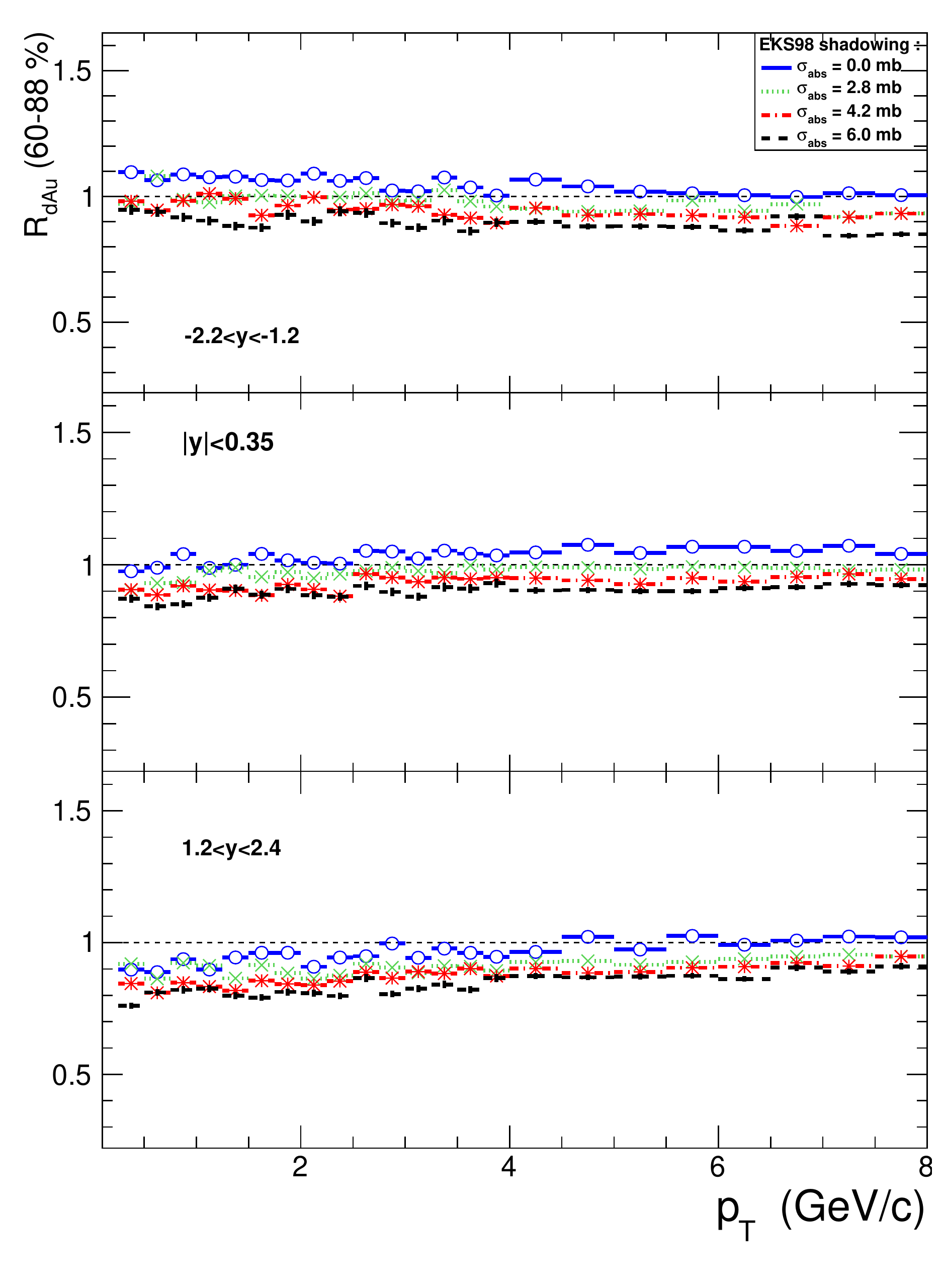}}
\subfloat[EPS08]{\includegraphics[width=0.33\textwidth,keepaspectratio]{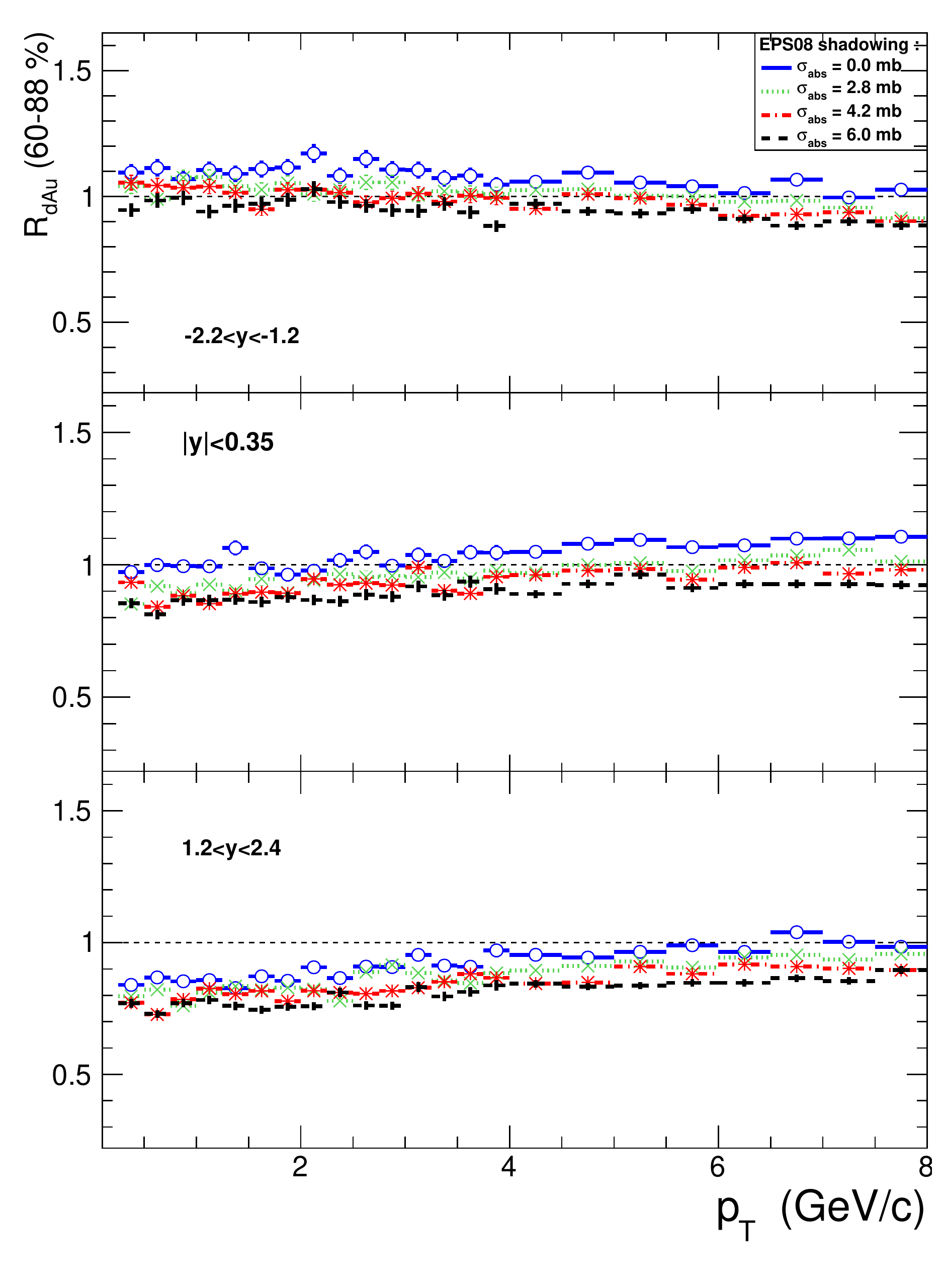}}
\subfloat[nDSg]{\includegraphics[width=0.33\textwidth,keepaspectratio]{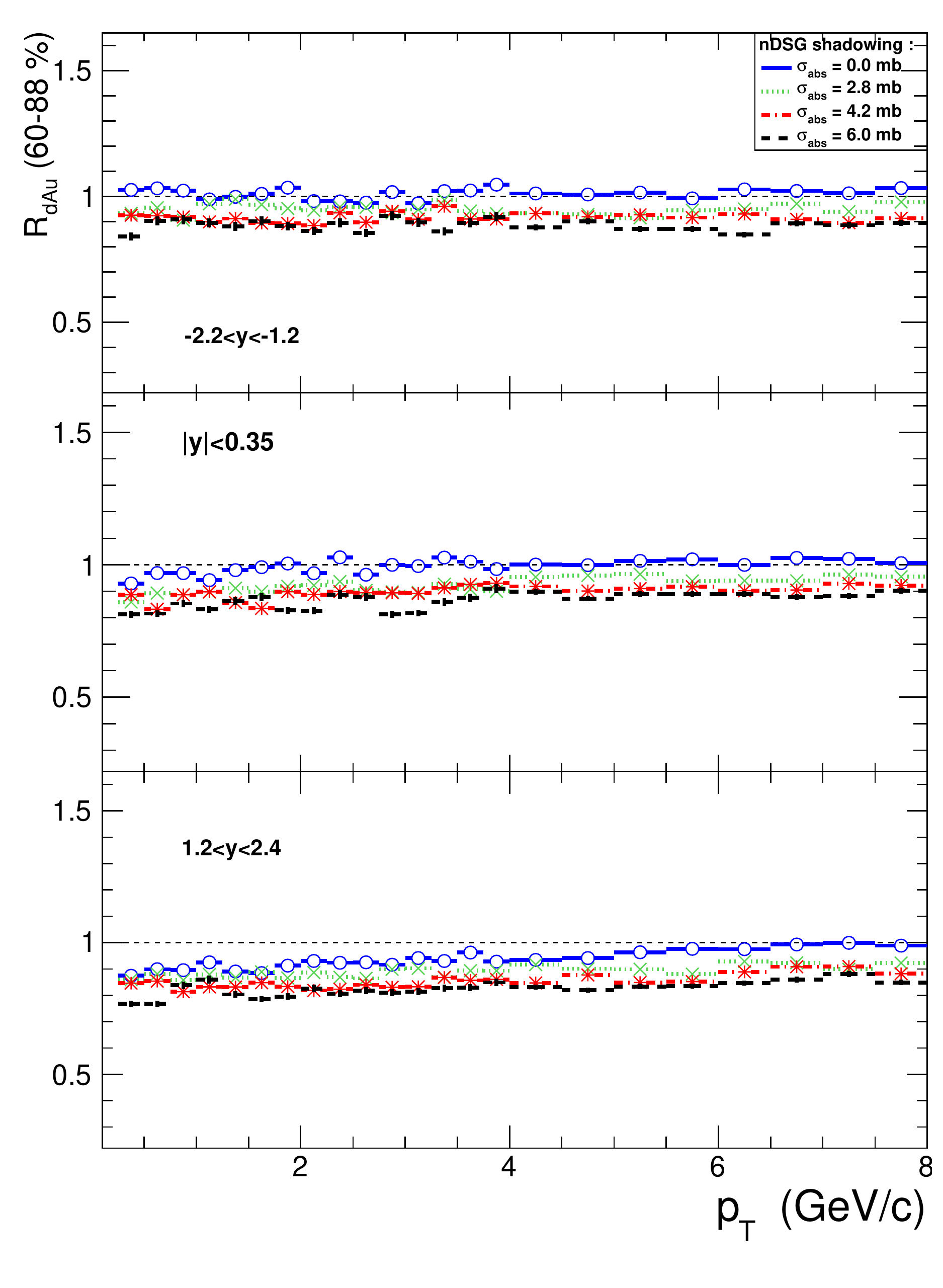}}
\caption{Idem as the Fig.~\protect\ref{R_dAu_flatpT_3_y_ranges} for the centrality class 60-88 \%.}
\end{center}
\end{figure}

\end{document}